\documentclass[aps,reprint,twocolumn,superscriptaddress,citeautoscript,amsmath,amssymb,floatfix,longbibliography,prx,]{revtex4-2}
\usepackage{amsmath}
\usepackage{amssymb}
\usepackage{amsfonts}
\usepackage{graphicx}
\usepackage{siunitx}
\usepackage{bm}
\usepackage{gensymb}
\usepackage{xcolor}
\usepackage{hyperref}

\begin{document}

\title[]{2D coherent spectroscopy of topological insulators}

\author{Tha\'{i}s V. Trevisan}
\affiliation {S\~{a}o Carlos Institute of Physics, University of S\~{a}o Paulo, 13560-970, S\~{a}o Carlos, SP, Brazil}
\author{Micha{\l} Papaj}
\affiliation {Department of Physics and Texas Center for Superconductivity (TcSUH) at the University of Houston, Houston,
Texas 77204, USA}

\begin{abstract}
Multidimensional coherent spectroscopy probes the response of a system to a sequence of time-separated light pulses, and has attracted significant interest in recent years. In the frequency space, it generates multidimensional spectra, in which information about different degrees of freedom of the system is generically disentangled. In this work, we develop a quantum geometric formulation of multidimensional optical susceptibilities in insulating systems. Focusing on a two-pulse setup and working in the velocity gauge, we derive analytical expressions for second-order and third-order two-dimensional coherent spectra (2DCS). We focus on the third-order ABB response, which remains finite in centrosymmetric systems where the second-order susceptibility vanishes by inversion symmetry. Its paramagnetic contribution supports an echo signal that probes products of four interband Berry-connection matrix elements, which reduce in two-band systems to bilinear combinations of the quantum geometric tensor of the occupied band. Using continuum and square-lattice models of two-dimensional Chern insulators, we show how the echo resolves the momentum-space sign structure of these geometric quantities. For the $xxyx$ response of the models considered, the number of nodes in the echo signal is related to the band topology. Our results extend the quantum-geometric interpretation of multidimensional spectroscopy to higher-order responses, showing that third-order 2DCS provides access to geometric structures beyond those probed at lower orders.
\end{abstract}

\date{\today}
		
\maketitle

\section{Introduction}

In the years that followed the discovery of the quantum Hall effect~\cite{klitzing1980new}, a paradigm shift in the understanding of the electronic properties of solids took place. Many theoretical and experimental advancements showed that a complete description of the electronic properties of solids cannot be achieved solely in terms of features of the electronic bands, such as band gaps and band curvatures. Instead, it is crucial to consider information encoded in the electronic wave function. For instance, the Berry phase~\cite{berry1984quantal}, a geometrical phase accumulated along the evolution of the state through a path in the Brillouin zone or in the parameter space of the Hamiltonian, is fundamental to the description of polarization~\cite{resta2007theory} and magnetization~\cite{thonhauser2011theory,vanderbilt2018berry} in solids. More broadly, the geometry of the electronic eigenstate space, called quantum geometry, deeply affects multiple physical properties of solids~\cite{torma2023essay,torma2022superconductivity,yu2025quantum,ahn2020low,ahn2022riemannian}.  

The concept of quantum geometry associated with the state space of a quantum system was introduced in the 80s in the context of clean electronic systems in which interactions are weak and can be neglected. In this case, the quantum geometry is formulated in terms of a complex tensor called the quantum geometric tensor (QGT), which is calculated in terms of the electronic Bloch wave functions~\cite{provost1980riemannian,shapere1989geometric} (see also Sec.\ref{sec:connection_geometry} and Sec.\ref{subsec:two_bands}).  The real part of the QGT is the quantum metric tensor, a symmetric tensor that provides a measure of infinitesimal distance between pairs of Bloch states. The imaginary part of the QGT is proportional to the Berry curvature, which is related to the phase acquired by a state upon being transported on a loop in the Bloch sphere~\cite{torma2023essay,yu2025quantum}. 

The integral of the Berry curvature in the Brillouin zone characterizes the topology of the space of the Bloch states. For instance, in a two-dimensional insulator, the integral of the Berry curvature over the first Brillouin zone (BZ) is quantized in units of $2\pi$, defining the Chern number $C$~\cite{vanderbilt2018berry,xiao2010berry,BernevigHughes2013}. Insulating states with finite $C$ host chiral metallic states localized at the boundaries of the crystal, resulting in a quantized anomalous linear Hall response in these systems~\cite{thouless1982quantized,kohmoto1985topological}. Moreover, insulating states with distinct values of $C$ cannot be smoothly connected without closing the bulk gap and, therefore, belong to distinct topological classes. Two-dimensional insulators with a nonzero Chern number are commonly referred to as Chern insulators~\cite{haldane1988model,qi2011topological,BernevigHughes2013}. More generically, the Berry curvature gives an anomalous component to the group velocity of electrons in solids, tied to the origins of the anomalous Hall effect~\cite{nagaosa2010anomalous}. Besides, higher-order moments of the Berry curvature can be generically finite in time-reversal even systems, and were shown to originate nonlinear Hall responses~\cite{sodemann2015quantum}.

\begin{figure}[t!]
    \centering
    \includegraphics[width=0.99\linewidth]{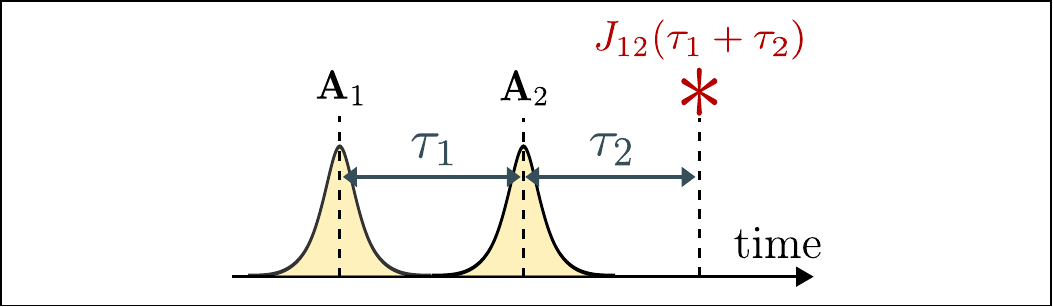}
    \caption{Illustration of the simplest light pulse setup for multidimensional spectroscopy. Two sharp pulses with vector potential $\mathbf{A}_1$ and $\mathbf{A}_2$ interact with a system at times separated by the interval $\tau_1$. A coherent electromagnetic signal is generated by the system which is measured at a time $\tau_1+\tau_2$ and related to the current density induced by the incoming pulses. Such outgoing signal carries information on the microscopic degrees of freedom of the system.}
    \label{fig:setup_2p}
\end{figure}

The significance of the quantum metric has recently started to attract more attention. For instance, it was shown that the quantum metric also contributes to the electronic group velocity, leading to an intrinsic nonlinear Hall response that has a dominant contribution in systems invariant under the product of spatial inversion and time reversal symmetries~\cite{gao2014field,gao2023quantum}. The search for observable manifestations of quantum geometry has rapidly expanded from its paradigmatic role in Hall-type transport and is becoming an organizing principle across optical, transport, and superconducting phenomena~\cite{yu2025quantum}. In nonlinear optics, it was shown that the shift-current contribution to the bulk photovoltaic effect admits a geometric description~\cite{ahn2020low}. More broadly, resonant optical processes have been expressed in a Riemannian-geometric language of transition dipole matrix elements~\cite{ahn2022riemannian}. Additionally, recent work also unveiled geometric interpretation of the giant magnetoresistance~\cite{neto2025gate} and magnetochiral anisotropy~\cite{fontana2025quantum}. In superconductors, the quantum metric contributes to the superfluid weight of multiband systems and can play an essential role in flat-band superconductivity~\cite{huhtinen2022revisiting,torma2022superconductivity}.

In this work, we address the quantum geometric interpretation of the multidimensional optical susceptibilities of solids. The multidimensional spectrum corresponds to the response of a system to a sequence of time-spaced light pulses~\cite{li2023optical,cundiff2013optical,mukamel1995principles}. The simplest setup is illustrated in Fig.~\ref{fig:setup_2p} and consists of two ultrafast light pulses incident on matter in times separated by an interval $\tau_1$. Outgoing electromagnetic signal generated by matter is measured after a subsequent time interval $\tau_2$, from which the system's response can be extracted. Importantly, $\tau_1$ and $\tau_2$ are independent, yielding a two-dimensional (2D) response in time-domain. The spectroscopic features can be more readily interpreted in the frequency domain and thus a Fourier transform can be performed with respect to both intervals $\tau_1$ and $\tau_2$, yielding a two-dimensional response in the frequency-domain called the 2D coherent spectrum (2DCS). More broadly, incoming light trains with more than two pulses lead to higher-dimensional spectra, generically called multidimensional coherent spectroscopy (MCS)~\cite{tollerud2017coherent}. 

MCS was developed in the context of nuclear magnetic resonance~\cite{ernst1990principles} and it was later implemented in the optical regime~\cite{tanimura1993two,golonzka2001vibrational,hamm1999two}, where it was demonstrated to provide clear spectroscopic signatures of molecular vibrational modes~\cite{khalil2004vibrational} and dynamics of chemical reactions~\cite{zheng2005ultrafast}. With the advent of THz ultrafast light pulses, MCS started to be applied to many-body electronic systems~\cite{lu2018two} and has been acquiring growing interest from the condensed matter community in the past few years~\cite{nandkishore2021spectroscopic,wan2019resolving,lu2017coherent,liu2020spectroscopic,qiang2024probing} due to its potential to unveil more information about the microscopic degrees of freedom of a many-body system and their dynamics in comparison with other one-dimensional (linear and non-linear) spectroscopic techniques. This is because the interaction with light pulses at different time instances leads to correlations between spectral features originating from different degrees of freedom and quasi-particles of the system, making it possible to disentangle their signatures~\cite{li2023optical}. MCS can also resolve the homogeneous and inhomogeneous broadening of ensembles of particles, revealing information on quasi-particle lifetimes often hidden in linear response~\cite{wan2019resolving}.    

Applications of MCS in condensed matter physics are rapidly developing. Recent work has begun to establish two-dimensional coherent spectroscopy as a probe of quantum geometry, showing that the two-frequency second-order nonlinear conductivity decomposes into distinct geometric contributions~\cite{froese2026probing}. Here, we develop a complementary formulation in terms of multidimensional optical susceptibility tensors of insulating solids within the velocity gauge.

We derive analytic expressions for second-order and third-order 2DCS in terms of combinations of components of the non-Abelian Berry connection. In the two-band limit, these combinations reduce to the quantum geometric tensor and higher-order combinations built from it. We focus primarily on the third-order 2DCS, which is symmetry allowed in centrosymmetric systems where the second-order 2DCS is forced to vanish due to spatial inversion symmetry. The same procedure, however, can be extended to higher-order multidimensional responses.

We show that spectral features in different regions of the frequency plane encode distinct quantum-geometric quantities. In particular, for the third-order ABB 2DCS discussed in detail in Sec.~\ref{sec:MDCS_theory}, the features along the anti-diagonal direction, $\omega_2=-\omega_1$, give rise to an echo signal sensitive to a product of four non-Abelian Berry-connection matrix elements. To illustrate these results, we compute the ABB spectra of minimal models of Chern insulators and show that their spectral features can also distinguish different topological regimes.

This manuscript is organized as follows. In Sec.~\ref{sec:MDCS_theory}, we develop the geometric theory of multidimensional optical susceptibility tensors, which constitutes the main result of this work. We first introduce the formalism of multidimensional spectroscopy and the associated multi-time optical correlation functions in Sec.~\ref{sec:suscep}. Next, in Sec.~\ref{sec:connection_geometry}, we connect these correlation functions to the band geometry of a generic multiband insulator and show that multidimensional spectra contain geometric quantities beyond those commonly considered in conventional single-frequency nonlinear optical responses. In Sec.~\ref{subsec:two_bands}, we focus on the two-band limit of the theory, where a direct connection with the quantum geometric tensor of the valence band can be established. In Sec.~\ref{sec:results}, we present the theoretical 2DCS of two minimal Chern-insulator models and show that the spectral features of the third-order ABB response are sensitive not only to band geometry, but also to band topology. We emphasize that the formalism we develop here can be readily applied to a broad class of insulating systems.

\section{Theory of multidimensional spectroscopy}\label{sec:MDCS_theory}

In this section, we present the formalism of multidimensional spectroscopy applied to many-body electronic systems. We focus on weakly correlated regimes in which an effective single-particle tight-binding description provides an adequate starting point. We derive analytical expressions for the multidimensional optical susceptibility tensors in the velocity gauge~\cite{parker2019diagrammatic,ventura2017gauge,passos2018nonlinear} and provide a quantum geometric interpretation for these responses. The derivation uses standard response-function and density-matrix approaches widely used in multidimensional coherent spectroscopy~\cite{mukamel1995principles,li2023optical}. We make an explicit connection between multidimensional responses and the quantum geometry of the Bloch states, which is the central result of this work. 

The calculation of the multidimensional optical susceptibility tensors can also be adapted to interacting systems. As an example, two-dimensional coherent spectroscopy has been theoretically formulated for exactly solvable quantum magnets such as the Kitaev honeycomb spin liquid~\cite{nandkishore2021spectroscopic,qiang2024probing}. Extensions to interacting systems in which the many-body spectrum is not available in closed form have also recently been developed~\cite{chen2025multidimensional,valmispild2024sub}.

%----

\subsection{Multidimensional optical susceptibilities}\label{sec:suscep}

The multidimensional spectrum is an intrinsically nonlinear response of the system to irradiation with a sequence of time-ordered light pulses. The simplest setup consists of two pulses, as illustrated in Fig.~\ref{fig:setup_2p}. The system interacts first with pulse 1, and, after a time interval $\tau_1$, with pulse 2. As a result, a time-dependent polarization is induced in the system that generates an outgoing signal measured after a time $\tau_2$ has passed since the interaction with pulse 2. The light-induced polarization can be further connected to other observables~\cite{mukamel1995principles}, such as the current density, which we focus on in this work. For concreteness, we denote by $\mathbf{J}_{12}(t)$ the time-dependent current density induced by the pulse sequence. Treating the light-matter interaction within perturbation theory, $\mathbf{J}_{12}(t)$ can be expanded in powers of the vector potential of the light field $\mathbf{A}(t)=\mathbf{A}_1(t)+\mathbf{A}_2(t)$, where $\mathbf{A}_i(t)$ denotes the vector potential of pulse $i$ (with $i=1,2$). The expansion is as follows~\cite{mukamel1995principles,watanabe2020general}      
\begin{widetext}
\begin{equation}
    J_{12,\mu}(t)=\int\limits_{-\infty}^{\infty}ds \,\chi_{\mu\nu}^{(1)}(s)A_{\nu}(t-s)+\int\limits_{-\infty}^{\infty}ds_1\int\limits_{-\infty}^{\infty}ds_2 
    \,\chi_{\mu\nu\gamma}^{(2)}(s_1,s_2)A_{\nu}(t-s_1)A_{\gamma}(t-s_2)+\mathcal{O}(A^3) \text{ ,}
    \label{eq:J_full}
\end{equation}
\end{widetext}

\noindent where $\mathcal{O}(A^3)$ denotes higher-order terms, and Einstein notation is implicit. The response functions $\chi^{(n)}(s_1,\cdots,s_n)$ are the multi-time optical susceptibilities from which the multidimensional spectra are obtained, as we discuss in detail next. Throughout this manuscript, parentheses in superscripts are used solely to indicate the order of the response and do not denote symmetrization over tensor indices. In Eq.~\eqref{eq:J_full} and throughout this work, we use the dipole approximation, in which the light wave vector is set to zero ($q=0$), which is justified by $q$ being much smaller than typical values of Fermi momenta. Within this approximation, the spatial dependence of current density, vector potential, and optical susceptibilities can be neglected. 

Furthermore, we use the velocity gauge, so that the coupling between a system and light is described by the Peierls substitution. An expansion of the resulting light-matter Hamiltonian in powers of $\mathbf{A}(t)$ yields two types of contributions to $\chi^{(n)}(s_1,\cdots,s_n)$: the paramagnetic contribution, or single-photon term, and the multi-photon terms, or contact terms, which we discuss in more detail next (see also Appendix \ref{app:chi_J}). Importantly, because of the time-ordering of the pulses in multidimensional spectroscopy, the multi-photon terms only contribute to the multidimensional spectra with $n\geq 3$. We return to this point later in this section and we also refer the reader to the Appendix \ref{app:chi_J} for details. 

\begin{figure}[b!]
    \centering
    \includegraphics[width=0.9\linewidth]{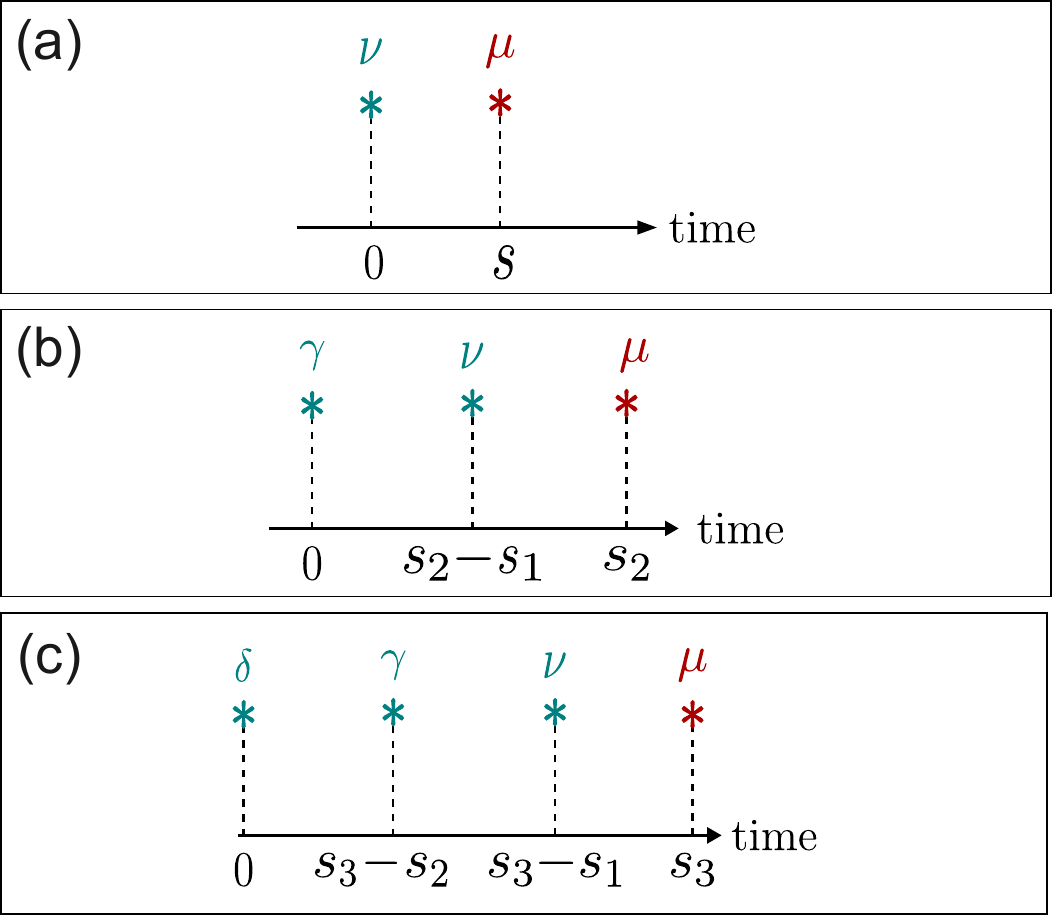}
    \caption{Illustration of the structure of the $n$-th order optical susceptibility tensors with (a) $n=1$ (linear response), (b) $n=2$ and (c) $n=3$. The green stars represent the interaction with the light field at the corresponding time instances. The red stars represent the measurement. The Greek letters associated to the stars are the Cartesian components of the current operator involved in the Kubo formula.}
    \label{fig:structure}
\end{figure}

The first term in Eq.~\eqref{eq:J_full} dominates when the applied light field is weak compared to the internal fields of the material. Here, $\chi_{\mu\nu}^{(1)}(s)$ denotes the linear (first-order) optical susceptibility tensor. As illustrated in Fig.~\ref{fig:structure}(a), it quantifies the contribution to the current at time $t$ arising from the interaction with the field at the earlier time $t-s$. Due to time-translation invariance, the linear response tensor depends only on the time-interval $s$ between the perturbation and the measurement of the response, with $s>0$ imposed by causality. For insulating systems, the optical responses have only interband contributions which, within the dipole approximation, account for vertical optical transitions between pairs of bands. Such an interband susceptibility is given by the standard Kubo formula~\cite{bruus2004many}, 
\begin{equation}
    \chi_{\mu\nu}^{(1)}(s)=\frac{i}{V}\theta(s)\left\langle \left[\hat{J}_{0,\mu}(s),\hat{J}_{0,\nu}(0)\right]\right\rangle \text{ .}
    \label{eq:chi_1_t}
\end{equation}

\noindent Here, $V$ denotes the volume of the system, the Heaviside step function $\theta(s)$ guarantees the causality of the response. Besides, $\hat{J}_{0,\mu}(t)=e^{i\hat{H}_0t}\hat{J}_{0,\mu}e^{-i\hat{H}_0t}$ denotes the Cartesian component of the single-photon current vertex (with $\mu=x,y,z$) in the interaction picture. The single-photon current vertex is the lowest-order term in the series expansion of the current operator in powers of the vector field of the light~\cite{BernevigHughes2013,watanabe2020general,parker2019diagrammatic}, and we discuss its expression in more detail in Sec.\ref{sec:connection_geometry} (see also Appendix \ref{app:chi_J}).  Moreover, $\hat{H}_0$ denotes the many-body Hamiltonian of the system. We keep the discussion generic in this section, while specific expressions for $\hat{H}_0$ and $\hat{J}_{0,\mu}$ in the context of Chern insulators are provided in Sec.\ref{sec:results}. Hereafter, we reserve the hat symbol ``\^{}'' to denote operators to avoid confusion between the current operator and the macroscopic current density $J_{\mu}(t)$ defined in Eq.(\ref{eq:J_full}). We also focus on the zero-temperature limit, such that $\left\langle\cdots\right\rangle$ denotes the expectation value in the ground state of $\hat{H}_0$. 

The Fourier transform of Eq.(\ref{eq:chi_1_t}),
\begin{equation}
\chi_{\mu\nu}^{(1)}(\omega)=\int\limits_{-\infty}^{\infty}ds\,\chi_{\mu\nu}^{(1)}(s)e^{i(\omega+i\Gamma)s} \text{ ,}  
\label{eq:chi1_freq}
\end{equation}

\noindent yields the linear spectrum of the system in frequency domain. Here, the parameter $\Gamma\rightarrow 0^{+}$ is a small real number introduced for the convergence of the integral, and can be interpreted as a phenomenological description of the decoherence and dephasing between the excitations promoted by the light field~\cite{li2023optical}.

We emphasize that Eq.(\ref{eq:chi_1_t}) contains only the paramagnetic contribution to the linear response. The contact term in first-order perturbation theory originates the Drude weight, which does not contribute to the zero-temperature response of insulators. As shown in the Appendix \ref{app:chi_J}, the contact term also yields a frequency-independent Fermi sea interband contribution for the linear response, which is generically finite in insulators. To emphasize the frequency-dependent resonant features in the results shown in Sec.\ref{sec:results}, we subtract this constant contact contribution from the displayed linear-response spectra.

For stronger incoming fields (but not strong enough to modify the equilibrium energy levels of the system and break the validity of perturbation theory), higher-order terms in the right-hand side of Eq.(\ref{eq:J_full}) become important. The function $\chi^{(n)}(s_1,\cdots,s_n)$, with $n>1$, is a multi-time $(n+1)$-rank tensor that quantifies the contribution to the current at time $t$ arising from $n$ interactions with the light field at earlier times $t-s_j$ with $j=1,2,\cdots n$, and $s_j>0$ as illustrated in Fig.\ref{fig:structure}. We show next that their Fourier transforms are related to the multidimensional spectra of the system. 

Importantly, to obtain the multidimensional spectrum, it is necessary to subtract from $\mathbf{J}_{12}(t)$ the current densities $\mathbf{J}_{1}(t)$ and $\mathbf{J}_2(t)$ induced by each pulse individually. This procedure removes linear terms in $\mathbf{A}(t)$, yielding an intrinsically non-linear current $\mathbf{J}_{NL}(t)=\mathbf{J}_{12}(t)-\mathbf{J}_{1}(t)-\mathbf{J}_{2}(t)$. Note that this subtraction also cancels non-linear interactions with a single pulse of the sequence. That is to say that $\mathbf{J}_{NL}(t)$ depends only on cross-interactions with the pulses of the sequence. This is an important difference between the multidimensional spectrum and other nonlinear responses such as high harmonic generation and the shift current, as the latter comes from nonlinear interaction with a \textit{single} light pulse. 

The discussion presented so far relies on the time interval between the pulses to be well defined. In other words, the duration of each pulse in the sequence should be short enough in comparison with their time intervals to avoid pulse overlap. To fulfill this condition, we focus on electromagnetic pulses much shorter than any characteristic time scale of the system, which we model by Dirac-delta pulses, following Ref.~\onlinecite{wan2019resolving}:
\begin{equation}
    \mathbf{A}(t)=\mathbf{A}_{1}\delta(t)+\mathbf{A}_{2}\delta(t-\tau_1) \text{ .}
    \label{eq:A_sequence}
\end{equation}

\noindent Here, $\mathbf{A}_{i}$, with $i=1,2$, are real vectors. Their magnitudes set the strengths of the corresponding impulsive vector-potential perturbations, while their directions define the linear polarizations of the corresponding pulses. More realistic pulses with finite width can be implemented \cite{li2023optical}, but would not change the qualitative features in our results shown in Sec.\ref{sec:results}. 

Substituting Eq.(\ref{eq:A_sequence}) into Eq.(\ref{eq:J_full}) and calculating the non-linear current at time $\tau_1+\tau_2$, $\mathbf{J}_{NL}(\tau_1+\tau_2)$, we obtain, up to third order in $A_{0,i}$, the following current density: $J_{NL,\mu}(\tau_1+\tau_2)=J_{NL,\mu}^{(2)}(\tau_1+\tau_2)+J_{NL,\mu}^{(3)}(\tau_1+\tau_2)$, with
\begin{equation}
    J_{NL,\mu}^{(2)}(\tau_1+\tau_2)=\chi_{\mu\nu\gamma}^{(2)}(\tau_2,\tau_1+\tau_2)A_{2,\nu}A_{1,\gamma} \text{ ,}
    \label{eq:J_NL2}
\end{equation}
\begin{align}
   J_{NL,\mu}^{(3)}(\tau_1+\tau_2)&=\chi_{\mu\nu\gamma\delta}^{(3)}(\tau_2,\tau_1+\tau_2,\tau_1+\tau_2) A_{2,\nu}A_{1,\gamma}A_{1,\delta}\nonumber\\[0.2cm]
    &+\chi_{\mu\nu\gamma\delta}^{(3)}(\tau_2,\tau_2,\tau_1+\tau_2) A_{2,\nu}A_{2,\gamma}A_{1,\delta} \text{ ,}
    \label{eq:J_NL3}
\end{align}

\noindent Here, the susceptibility tensor
\begin{align}
    &\chi_{\mu\nu\gamma}^{(2)}(s_1,s_2)=\nonumber\\[0.2cm]
    &\frac{i^2}{V}\theta(s_1)\theta(s_2-s_1)\left\langle\left[\left[\hat{J}_{0,\mu}(s_2),\hat{J}_{0,\nu}(s_2-s_1)\right],\hat{J}_{0,\gamma}(0)\right]\right\rangle \text{ .}
    \label{eq:chi_t_2}
\end{align}

\noindent is obtained in second-order perturbation theory in $\mathbf{A}(t)$\cite{watanabe2020general,mukamel1995principles}, and corresponds to the lowest-order response tensor that accounts for the interaction of the system with both pulses of the sequence. Note that causality requires $s_2>s_1$~\cite{boyd2008nonlinear}. The time ordering of the pulses in the sequence requires that the interaction happens first with pulse $1$ and later with pulse $2$, which constrains $s_2=\tau_1+\tau_2$ and $s_1=\tau_2$. Since $\tau_1$ and $\tau_2$ are independent time intervals, the Fourier transform
\begin{align}
    &\chi_{\mu\nu\gamma}^{(2)}(\omega_1,\omega_2)=\nonumber\\
    &=\int d\tau_1 d\tau_2 \,\chi_{\mu\nu\gamma}^{(2)}(\tau_2,\tau_1+\tau_2)\,e^{i\left(\omega_1+i\Gamma\right)\tau_1}e^{i\left(\omega_2+i\Gamma\right)\tau_2} \text{ ,}
    \label{eq:2DCS_freq}
\end{align}

\noindent yields a function of two independent frequencies $\omega_1$ and $\omega_2$, which is the so-called 2D coherent spectrum of the system (2DCS)~\cite{mukamel1995principles,li2023optical}. Here, $\Gamma$ has the same meaning as in Eq.(\ref{eq:chi1_freq}). We emphasize that the contact terms that contribute to the second-order response of a system to a single light-pulse, such as the SHG and the shift current, are eliminated in the subtraction that yields $\mathbf{J}_{NL}(\tau_1+\tau_2)$. Thus, the 2DCS has contribution solely from the single-photon terms (see Appendix \ref{app:chi_J}). Moreover, unlike single-pulse second-order responses, $\omega_1$ and $\omega_2$ are not the frequencies of the light pulses, but those conjugated to time intervals $\tau_1$ and $\tau_2$. 

\begin{figure}[t!]
    \centering
    \includegraphics[width=0.99\linewidth]{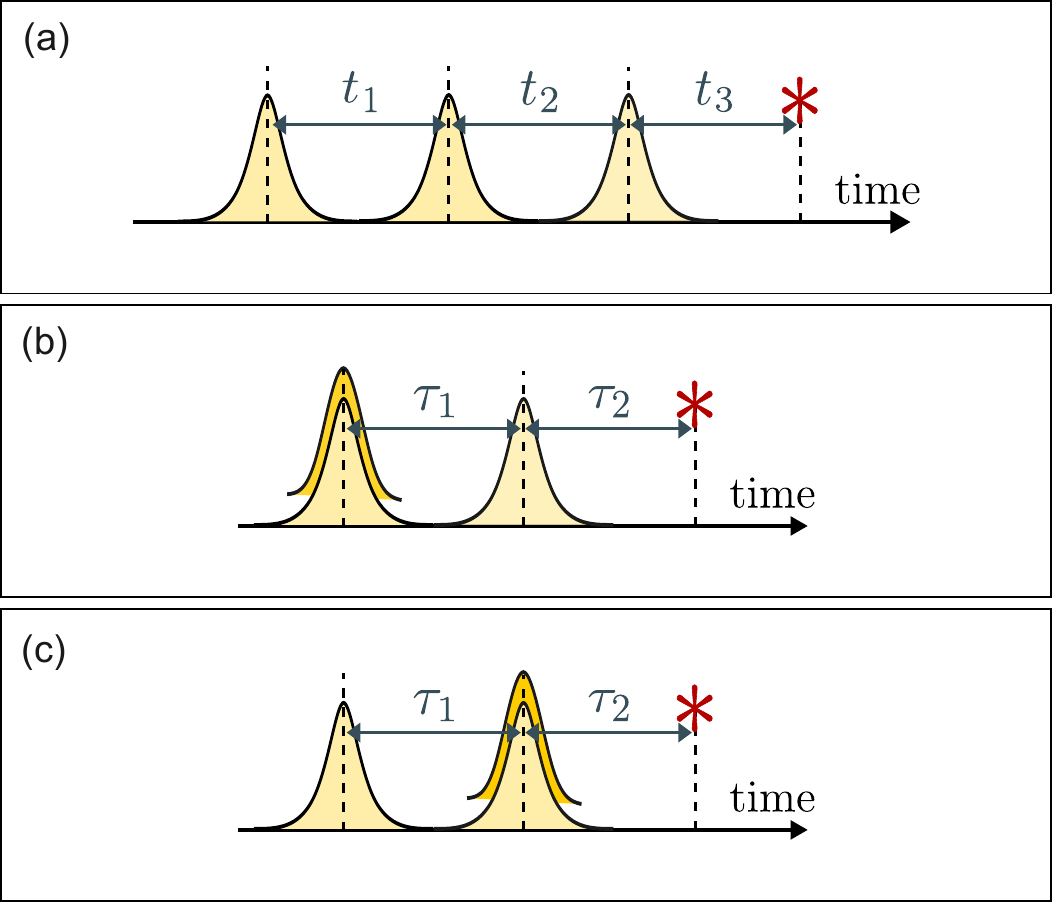}
    \caption{Illustration of the pulse setups for the third-order multidimensional responses. The three-pulse setup in (a) yields the 3DCS $\chi^{(3)}(t_3,t_2+t_3,t_1+t_2+t_3)$. The two-pulse setups in (b) and (c) give the third-order 2DCS called (b) AAB and (c) ABB. In the AAB response, the system interacts quadratically with pulse 1 and linearly with pulse 2 and the corresponding susceptibility tensor is obtained from (a) by setting $t_1=0$, $t_2=\tau_1$ and $t_3=\tau_2$. In the ABB response, the system interacts linearly with pulse 1 and quadratically with pulse 2 and the corresponding susceptibility tensor is obtained from (a) by setting $t_2=0$, $t_1=\tau_1$ and $t_3=\tau_2$.}
    \label{fig:pulse_setup}
\end{figure}

Moving to the next order in perturbation theory, the third-order response tensor $\chi_{\mu\nu\gamma\delta}^{(3)}(s_1,s_2,s_3)$, is the sum of a single-photon term 
\begin{align}    
&\chi_{\mu\nu\gamma\delta,\text{S}}^{(3)}(s_1,s_2,s_3)=\frac{i^3}{V}\theta(s_1)\theta(s_3-s_2)\theta(s_2-s_1)\times\nonumber\\[0.2cm]
    &\left\langle\left[\left[\left[\hat{J}_{0,\mu}(s_3),\hat{J}_{0,\nu}(s_3-s_1)\right],\hat{J}_{0,\gamma}(s_3-s_2)\right],\hat{J}_{0,\delta}(0)\right]\right\rangle \text{ ,}
    \label{eq:chi_3_t}
\end{align}

\noindent and multi-photon contact terms whose full expressions are provided in Appendix \ref{app:chi_J}. Below, we present the contact terms that contribute to the specific third-order 2DCS that we consider in this work.

The susceptibility $\chi_{\mu\nu\gamma\delta}^{(3)}(s_1,s_2,s_3)$ accounts for the interaction with light field at times $t-s_1$, $t-s_2$, $t-s_3$, resulting in a current density at $t$, as illustrated in Fig.\ref{fig:structure}(c). For the two-pulse setup considered here, two cases are possible: either all three interactions come from separate pulses, see Fig.~\ref{fig:pulse_setup}(a), or from two pulses with one of the pulses interacting quadratically, see Fig.~\ref{fig:pulse_setup}(b,c). The first yields a three-dimensional coherent spectra (3DCS) and, in this case, only the single-photon term contributes, as shown in Appendix \ref{app:chi_J}.  In this work, we focus on the second case, as we consider sequences with two pulses, relevant for 2D spectra. This scenario allows two possibilities for the third order contribution to the 2DCS. First, the system can interact quadratically with the pulse $1$ and linearly with pulse $2$, as illustrated in Fig.\ref{fig:pulse_setup}(b), and called AAB signal. The AAB susceptibility has a purely single-photon contribution, captured by $\chi_{\mu\nu\gamma\delta,\text{S}}^{(3)}(\tau_2,\tau_1+\tau_2,\tau_1+\tau_2)$, and a mixed single-photon and two-photon contribution $\chi_{\mu\nu\gamma\delta,\text{T}}^{(\text{ABB})}(\tau_2,\tau_1+\tau_2)$, with,
\begin{align}
    &\chi_{\mu\nu\gamma\delta;\text{T}}^{(\text{AAB})}(s_1,s_2)=\nonumber\\[0.2cm]
    &\frac{i^2\theta(s_1)\theta(s_2-s_1)}{2V}\left\langle\left[\left[\hat{J}_{0,\mu}(s_2),\hat{J}_{0,\nu}(s_2-s_1)\right],\hat{J}_{1,\gamma\delta}(0)\right]\right\rangle \text{ .}
    \label{eq:chi_AABST}
\end{align}

\noindent Here, $\hat{J}_{1,\gamma\delta}$ is the two-photon vertex obtained from the first-order expansion of the total current operator in powers of $\mathbf{A}(t)$ [see also Eq.(\ref{eq:J1_orbital})]. The second possibility is that the system interacts linearly with pulse $1$ and quadratically with pulse $2$, which is called the ABB signal. It also has a purely paramagnetic term given by $\chi_{\mu\nu\gamma\delta,\text{S}}^{(3)}(\tau_2,\tau_2,\tau_1+\tau_2)$ and a mixed (or contact) term 
\begin{align}
    &\chi_{\mu\nu\gamma\delta;\text{T}}^{(\text{ABB})}(s_1,s_2)=\nonumber\\[0.2cm]
    &\frac{i^2\theta(s_1)\theta(s_2-s_1)}{2V}\left\langle\left[\left[\hat{J}_{0,\mu}(s_2),\hat{J}_{1,\nu\gamma}(s_2-s_1)\right],\hat{J}_{0,\delta}(0)\right]\right\rangle \text{ .}
    \label{eq:chi_ABBST}
\end{align}

\noindent The AAB and the ABB responses have distinct spectral features. In particular, the ABB 2DCS has an echo signal, similar to the spin echo in nuclear magnetic resonance~\cite{wan2019resolving}, which we discuss in Sec.\ref{sec:results}.  
 
The spectral features in the 2DCS consist of a set of peaks in the frequency plane $\omega_1\times \omega_2$ which is a map of the coherent light-matter interactions. The position, amplitude, shape, and relative sign of these peaks carry information about the microscopic structure of the system, its excitations, interaction between excitations and their dynamics, which might be difficult to access in conventional single-frequency linear and non-linear responses~\cite{tollerud2017coherent,wan2019resolving,qiang2024probing,nandkishore2021spectroscopic}. In Sec.\ref{sec:connection_geometry}, we show that they also map the geometry of the space of the Bloch states, with different regions of the frequency plane being dominated by different quantum geometric quantities beyond the standard quantum metric and Berry curvature.   

The steps presented so far can be readily generalized for incoming light with more pulses. For a sequence with $l$ pulses that are separated by time intervals $\tau_1,\cdots \tau_{l-1}$, the lowest-order $l$-dimensional optical susceptibility tensor in the time domain, $\chi^{(l)}(\tau_l,\tau_l+\tau_{l-1},\cdots,\tau_1+\cdots\tau_l)$, is given by the $l$-th order Kubo formula involving nested commutators of $l+1$ single-photon vertices $\hat{\mathbf{J}}_0$. Here, $\tau_l$ is the time interval between the last pulse and the measurement of the response. In the frequency space, this yields a $l$-dimensional spectrum ($lDCS$). The case with $l=3$ was considered in Ref.~\onlinecite{liu2020spectroscopic}, where it was shown that the 3DCS provides unique fingerprints of electronic interaction in silicon vacancy centers in diamond. In general, the interpretation of the spectroscopic features in the multidimensional spectra is challenging~\cite{li2023optical}, and the difficulties increase with increasing dimensionality of the spectra.

There is a variety of possible setups for an incoming light field with more than two pulses. One possibility is to perform the Fourier transform in a subset of the $\tau_j$ and investigate the time-evolution of the spectral features in the $nDCS$ ($n<l$) as a function of the times kept in the response. This analysis provides information on the dynamics of the microscopic degrees of freedom of matter. In particular, this technique was applied to study the course of chemical reactions~\cite{zheng2005ultrafast}. Here, we focus on the 2DCS response of insulating systems, which, as we show in the next section, can be formulated in terms of the quantum geometry of the cell-periodic Bloch wave functions. 

\subsection{Connection with band geometry}\label{sec:connection_geometry}

In this section, we provide a quantum geometric interpretation of the spectral features of the multidimensional optical susceptibilities of insulators. As a starting point, we consider a non-interacting electronic Hamiltonian $\hat{H}_0=\sum\limits_{\mathbf{k}}\hat{\psi}_{\mathbf{k}}^{\dag}\hat{h}(\mathbf{k})\hat{\psi}^{\null}_{\mathbf{k}}$, where $\hat{\psi}^{\dag}_{\mathbf{k}}$ ($\hat{\psi}^{\null}_{\mathbf{k}}$) are the vectors of creation (annihilation) operators written in the basis of the internal degrees of freedom of the system, such as spin, orbital and/or sublattice, which we generically refer to as \textit{orbital basis}. The first-quantized Hamiltonian $\hat{h}(\mathbf{k})$ has eigenvalues $\varepsilon_{n}(\mathbf{k})$, and eigenstates $\left|n(\mathbf{k})\right\rangle$, which correspond to the electronic bands and cell-periodic Bloch states, respectively. Here, $n$ denotes the band index. The discussion carried out in this section does not depend on the specific form of $\hat{h}(\mathbf{k})$ and, thus, will be kept generic. Specific expressions for $\hat{h}(\mathbf{k})$ are given in Sec.\ref{sec:results}, in the context of Chern insulators. 

In the presence of light, and adopting the velocity gauge~\cite{parker2019diagrammatic,ventura2017gauge,passos2018nonlinear}, the Hamiltonian is modified to $\hat{H}[\mathbf{A}]$ by Peierls substitution~\cite{BernevigHughes2013,bruus2004many} $\hat{h}(\mathbf{k})\rightarrow\hat{h}(\mathbf{k}+\frac{e}{\hbar c}\mathbf{A}(t))$. Here, $e>0$ is the fundamental charge and $c$ is the speed of light. The current operator is obtained from $\hat{H}[\mathbf{A}]$, by calculating  $\hat{J}_{\mu}=-c\frac{\partial H\left[\mathbf{A}\right]}{\partial A_{\mu}}$ ~\cite{bernevig2013topological}, and it can be written as a power series in $\mathbf{A}(t)$ (see Appendix \ref{app:chi_J}). Here, we consider the first two coefficients of this series, which originate the terms that will contribute to the multidimensional spectra that we consider in this work:
\begin{align}
&\hat{J}_{0,\mu}=-\frac{e}{\hbar}\sum\limits_{\mathbf{k}}\hat{\psi}_{\mathbf{k}}^{\dag}\left(\partial_{\mu}\hat{h}(\mathbf{k})\right)\hat{\psi}_{\mathbf{k}}^{\null} \text{ ,}\label{eq:J_orbital}\\[0.2cm]
&\hat{J}_{1,\mu\nu}=-\frac{e^2}{c\hbar^2}\sum\limits_{\mathbf{k}}\hat{\psi}_{\mathbf{k}}^{\dag}\left(\partial_{\mu}\partial_{\nu}\hat{h}(\mathbf{k})\right)\hat{\psi}_{\mathbf{k}}^{\null} \text{ .}
\label{eq:J1_orbital}
\end{align}

\noindent The paramagnetic part of the current operator, also called single-photon vertex~\cite{parker2019diagrammatic}, is given by Eq.(\ref{eq:J_orbital}), while the first order term in Eq.(\ref{eq:J1_orbital}) corresponds to the two-photon vertex~\cite{parker2019diagrammatic}.  We define $\partial_{\mu}\equiv \frac{\partial}{\partial k_{\mu}}$ ($\mu=x,y,z$). 

To calculate the multidimensional optical susceptibility tensors, it is more convenient to work in the band basis, in which Eq.(\ref{eq:J_orbital}) and Eq.(\ref{eq:J1_orbital}) become

\begin{align}
&\hat{J}_{0,\mu}=-\frac{e}{\hbar}\sum\limits_{\mathbf{k},n,m}j_{nm}^{(\mu)}(\mathbf{k})c_{\mathbf{k}n}^{\dag}
    c_{\mathbf{k}m}^{\null} \text{ ,}\label{eq:J_band}\\[0.2cm]
 &\hat{J}_{1,\mu\nu}=-\frac{e^2}{c\hbar^2}\sum\limits_{\mathbf{k},n,m}s_{nm}^{(\mu\nu)}(\mathbf{k})c_{\mathbf{k}n}^{\dag}
    c_{\mathbf{k}m}^{\null}    
    \label{eq:J1_band}
\end{align}

\noindent with matrix elements
\begin{align}
   &j_{nm}^{(\mu)}(\mathbf{k})= \left\langle n(\mathbf{k})\left|\partial_{\mu}\hat{h}(\mathbf{k})\right|m(\mathbf{k})\right\rangle\text{ ,}
   \label{eq:J_bandmatrix}\\
   &s_{nm}^{(\mu\nu)}(\mathbf{k})= \left\langle n(\mathbf{k})\left|\partial_{\mu}\partial_{\nu}\hat{h}(\mathbf{k})\right|m(\mathbf{k})\right\rangle=s_{nm}^{(\nu\mu)}(\mathbf{k})\text{ .}
\end{align}

\noindent Note that, in the dipole approximation $\left[j_{nm}^{(\mu)}(\mathbf{k})\right]^*=j_{mn}^{(\mu)}(\mathbf{k})$, $\left[s_{nm}^{(\mu\nu)}(\mathbf{k})\right]^*=s_{mn}^{(\mu\nu)}(\mathbf{k})$, where the symbol $*$ denotes complex conjugation. 

The connection between band geometry and linear optical responses is already well established in the literature~\cite{ahn2022riemannian,ghosh2024probing}. However, we present this case before moving to the multidimensional responses to set notation and to make the work self-contained. Substituting Eq.(\ref{eq:J_band}) into Eq.(\ref{eq:chi_1_t}), and taking the Fourier transform we obtain  
\begin{equation}
    \chi_{\mu\nu}^{(1)}(\omega)=\frac{1}{V}\sum\limits_{\mathbf{k},n,m}\frac{f_{nm}(\mathbf{k})j_{mn}^{(\mu)}(\mathbf{k})j_{nm}^{(\nu)}(\mathbf{k})}{\omega-\Delta_{nm}(\mathbf{k})+i\Gamma} \text{ .}
    \label{eq:chi_1_omega}
\end{equation}

\noindent Here, $\Delta_{nm}(\mathbf{k})\equiv \varepsilon_n(\mathbf{k})-\varepsilon_m(\mathbf{k})$ is the energy difference between bands $n$ and $m$ and  $f_{nm}(\mathbf{k})=f_{n}(\mathbf{k})-f_{m}(\mathbf{k})$, where $f_{n}(\mathbf{k})$ denotes the Fermi-Dirac function of band $n$. Hereafter, we also set $e=\hbar=c=1$. Note that, due to $f_{nm}(\mathbf{k})$, only the off-diagonal components of $\hat{\mathbf{J}}_0$ (in the band basis) contribute to the linear response. They encode information about the quantum geometry of the Bloch wave functions, which can be readily seen by explicitly applying the momentum derivatives to the Bloch states in Eq.(\ref{eq:J_band}):
\begin{equation}
       j_{nm}^{(\mu)}(\mathbf{k})=v_{n,\mu}(\mathbf{k})\delta_{n,m}+i\Delta_{nm}(\mathbf{k})\mathcal{A}_{nm}^{(\mu)}(\mathbf{k})\text{ .}
\label{eq:j_Bloch}
\end{equation}

\noindent Here, $\delta_{n,m}$ denotes the Kronecker delta and $v_{n,\mu}(\mathbf{k})\equiv\partial_{\mu}\varepsilon_{n}(\mathbf{k})$ is the band velocity. Besides, $\mathcal{A}_{nm}^{(\mu)}(\mathbf{k})=i\left\langle \left.n(\mathbf{k})\right|\partial_{\mu}m(\mathbf{k})\right\rangle$ is the non-Abelian Berry connection~\cite{vanderbilt2018berry}, that is associated with quantum geometric effects. Substituting Eq.(\ref{eq:j_Bloch}) into Eq.(\ref{eq:chi_1_omega}), we obtain
\begin{equation}
    \chi_{\mu\nu}^{(1)}(\omega)=
    \frac{1}{V}\sum\limits_{\mathbf{k},n,m}\frac{f_{nm}\left(\mathbf{k}\right)\Delta_{nm}^{2}(\mathbf{k})\mathcal{A}_{mn}^{(\mu)}(\mathbf{k})\mathcal{A}_{nm}^{(\nu)}(\mathbf{k})}{\omega-\Delta_{nm}(\mathbf{k})+i\Gamma} \text{ .}
    \label{eq:chi_1_omega_A}
\end{equation}  

The spectral features of $\chi_{\mu\nu}^{(1)}(\omega)$ are governed by the interband transition energies $\Delta_{nm}(\mathbf{k})$ and their corresponding optical matrix elements.  In particular, resonant transitions occur when $\omega=\Delta_{nm}(\mathbf{k})$, with a phenomenological broadening constant $\Gamma$. The momenta satisfying the resonance condition, together with the occupation constraints, define the region of momentum space in which interband transitions are energetically available at a given frequency. We refer to this as the optically active region (OAR), with the understanding that its contribution to a given response component also depends on the relevant optical matrix elements. The density of available transitions is characterized by the joint density of states (JDOS)~\cite{wooten2013optical} $\rho(\omega)=\sum\limits_{n,m}\int d^2k\,f_{mn}(\mathbf{k})\delta\left(\omega-\Delta_{nm}(\mathbf{k})\right)$ (see also  Sec.\ref{sec:results}). The linear response contains an additional, transition-dependent weight  $\Delta_{nm}^{2}(\mathbf{k})\mathcal{A}_{mn}^{(\mu)}(\mathbf{k})\mathcal{A}_{nm}^{(\nu)}(\mathbf{k})$. Here, the product of Berry connections is gauge invariant and defines the pair-resolved quantum geometric tensor, $\mathcal{Q}_{\mu\nu}^{m\rightarrow n}(\mathbf{k})\equiv \mathcal{A}_{mn}^{(\mu)}(\mathbf{k})\mathcal{A}_{nm}^{(\nu)}(\mathbf{k})$, which characterizes the quantum geometry associated with the optical transition between bands $m$ and $n$ \cite{ahn2022riemannian}. In two-band systems, $\mathcal{Q}^{n\rightarrow m}(\mathbf{k})$ becomes the standard quantum geometric tensor of the valence band, as we show in Sec.\ref{subsec:two_bands}. Consequently, the spectral weights of the linear response reflect both the JDOS and the quantum geometry of the resonantly connected states. Depending on the tensor component, geometric contributions from different regions of momentum space can enhance or partially cancel one another. Concrete examples are discussed in Sec.~\ref{sec:results}.
  
The derivation of the expressions for the multidimensional optical susceptibilities introduced in Sec.\ref{sec:suscep} follows similar steps. Below, we show the final results in terms of the non-Abelian Berry connection, and the responses written in terms of the matrix elements of the current operator are listed in the Appendix \ref{app:chi_J}. We first consider the second-order 2DCS defined in Eq.(\ref{eq:chi_t_2}). After some algebra, we obtain $\chi^{(2)}(\omega_1,\omega_2)=\chi_+^{(2)}(\omega_1,\omega_2)+\chi_-^{(2)}(\omega_1,\omega_2)$, with
\begin{align}
    &\chi^{(2)}_+(\omega_1,\omega_2)=-\sum\limits_{\substack{\mathbf{k},n,m\\\varepsilon_n(\mathbf{k})>\varepsilon_m(\mathbf{k})}}\frac{f_{mn}(\mathbf{k})}{V}\left[\Delta_{nm}^{2}(\mathbf{k})\bar{P}_{nm}\right.\nonumber\\[0.2cm]
    &-i\sum\limits_{l\neq m,n}\left.\Delta_{ml}(\mathbf{k})\Delta_{ln}(\mathbf{k})\Delta_{nm}(\mathbf{k})\bar{R}_{nml}\right] \text{ .}
    \label{eq:chi_2main}
\end{align}

\noindent and $\chi_-^{(2)}(\omega_1,\omega_2)=\left[\chi^{(2)}_+(-\omega_1,-\omega_2)\right]^*$. Here $\bar{P}_{nm}$ and $\bar{R}_{nml}$ denote rank-three tensors that are functions of $\mathbf{k}$, $\omega_1$ and $\omega_2$, hereafter omitted to shorten the notation. Their matrix elements are the following:
 \begin{align}
     \bar{P}_{nm,\mu\nu\gamma}&\equiv\frac{\left[\partial_{\nu}\Delta_{nm}(\mathbf{k})\right]\mathcal{Q}_{\mu\gamma}^{m\rightarrow n}(\mathbf{k})}{\left(\omega_1-\Delta_{nm}(\mathbf{k})+i\Gamma\right)\left(\omega_2-\Delta_{nm}(\mathbf{k})+i\Gamma\right)}\nonumber\\[0.2cm]
     &-\frac{\left[\partial_{\mu}\Delta_{nm}(\mathbf{k})\right]\mathcal{Q}_{\nu\gamma}^{m\rightarrow n}(\mathbf{k})}{\left(\omega_1-\Delta_{nm}(\mathbf{k})+i\Gamma\right)\left(\omega_2+i\Gamma\right)}\text{ ,}
     \label{eq:P}
 \end{align}
 
 \noindent and 
 \begin{align}
     \bar{R}_{nml,\mu\nu\gamma}&\equiv\frac{\mathcal{A}_{ml}^{(\mu)}(\mathbf{k})\mathcal{A}_{ln}^{(\nu)}(\mathbf{k})\mathcal{A}_{nm}^{(\gamma)}(\mathbf{k})}{\left(\omega_1-\Delta_{nm}(\mathbf{k})+i\Gamma\right)\left(\omega_2-\Delta_{lm}(\mathbf{k})+i\Gamma\right)}\nonumber\\[0.2cm]
     &-\frac{\mathcal{A}_{ln}^{(\mu)}(\mathbf{k})\mathcal{A}_{ml}^{(\nu)}(\mathbf{k})\mathcal{A}_{nm}^{(\gamma)}(\mathbf{k})}{\left(\omega_1-\Delta_{nm}(\mathbf{k})+i\Gamma\right)\left(\omega_2-\Delta_{nl}(\mathbf{k})+i\Gamma\right)} \text{ .}
     \label{eq:R}
 \end{align}
 
The spectral features of $\chi^{(2)}(\omega_1,\omega_2)$ are peaks on the $\omega_1\times\omega_2$ frequency plane that are a map of the coherent light-matter interaction. As in linear response, the peak positions are set by the band structure. Here, however, the peak amplitudes and relative signs manifest geometric quantities that are beyond the quantum metric and Berry curvature. We focus first on the spectroscopic signatures originating from the $\bar{P}_{nm}(\mathbf{k})$ term. As can be seen from Eq.(\ref{eq:P}), the amplitude of peaks along the main diagonal ($\omega_1=\omega_2$) and the horizontal axis ($\omega_2=0$) in the frequency plane probe different Cartesian components of the OAR-integrated tensor $\left[\partial_{\mu}\Delta_{nm}(\mathbf{k})\right]\mathcal{Q}_{\nu\gamma}^{m\rightarrow n}(\mathbf{k})$. This quantity is the relative-velocity-weighted quantum geometric tensor associated with the pair of bands $(n,m)$ and, in a two-band system, it is related to the frequency-resolved quantum metric (real part) and Berry curvature dipoles (imaginary part)\cite{sodemann2015quantum,wang2021intrinsic}.

The term $\bar{R}_{nml}(\mathbf{k})$ defined in Eq.(\ref{eq:R}) only contributes to $\chi^{(2)}(\omega_1,\omega_2)$ in systems with more than two bands and it leads to off-diagonal peaks in the spectra which probe different Cartesian components of the the product $\mathcal{A}_{nm}^{(\mu)}(\mathbf{k})\mathcal{A}_{ml}^{(\nu)}(\mathbf{k})\mathcal{A}_{ln}^{(\gamma)}(\mathbf{k})$. This is a three-band geometric contribution and cannot generally be reconstructed from the pair-resolved metric and curvature alone.

We consider now the third-order 2DCS, focusing on the ABB response, which has both single-photon [Eq.(\ref{eq:chi_3_t})] and two-photon contributions [Eq.(\ref{eq:chi_ABBST})]. In the frequency space, we obtain $\chi^{(\text{ABB})}_S(\omega_1,\omega_2)=\chi_{\text{S}+}^{(\text{ABB})}(\omega_1,\omega_2)+\chi_{\text{S}-}^{(\text{ABB})}(\omega_1,\omega_2)$ and $\chi_\text{T}^{(\text{ABB})}(\omega_1,\omega_2)=\chi_{_\text{T}+}^{(\text{ABB})}(\omega_1,\omega_2)+\chi_{\text{T}-}^{(\text{ABB})}(\omega_1,\omega_2)$, with
\begin{widetext}
\begin{align}
    &\chi^{(\text{ABB})}_{\text{S}+}(\omega_1,\omega_2)=\frac{i}{2V}\left.\sum\limits_{\mathbf{k},n>m}f_{mn}(\mathbf{k})\Delta_{nm}(\mathbf{k})\right[\Delta_{nm}(\mathbf{k})\bar{S}_{nm}\left.-i\sum\limits_{l\neq n,m}\Delta_{ml}(\mathbf{k})\Delta_{ln}(\mathbf{k})\bar{T}_{nml}+\sum\limits_{l>p} {}^{'}\bar{W}_{nmlp}\right] \text{ ,}
    \label{eq:chi_ABB}
\end{align}
\begin{align}
    &\chi^{(\text{ABB})}_{\text{T}+}(\omega_1,\omega_2)=-\frac{1}{2V}\left.\sum\limits_{\mathbf{k},n>m}f_{mn}(\mathbf{k})\Delta_{nm}(\mathbf{k})\right[\bar{U}_{nm}\left.-\sum\limits_{l\neq n,m}\bar{V}_{nml}\right] \text{ ,}
    \label{eq:chi_ABBT}
\end{align}
\end{widetext}

\noindent and $\chi^{(\text{ABB})}_{\alpha-}(\omega_1,\omega_2)=\left[\chi_{\alpha+}^{(\text{ABB})}(-\omega_1,-\omega_2)\right]^*$ ($\alpha=\text{S},\text{T}$; see Appendix \ref{app:chi_J}). In the previous equation, $\sum\limits_{l>p}{}'$ denotes the sum over the band indices $l$ and $p$, with $l>p$ and $(l,p)\neq(n,m)$. Besides, $\bar{S}_{nm}$, $\bar{T}_{nm}$,  $\bar{U}_{nm}$, $\bar{V}_{nm}$ and $\bar{W}_{nm}$ are rank-four tensors that are functions of $\mathbf{k}$, $\omega_1$ and $\omega_2$ (omitted to shorten notation) defined below:
\begin{widetext}
\begin{align}
\bar{S}_{nm,\mu\nu\gamma\delta}
&= \frac{\mathcal{A}_{mn}^{(\mu)}\mathcal{A}_{nm}^{(\delta)}\left[2\Delta_{nm}^{2}\mathcal{A}_{mn}^{(\gamma)}\mathcal{A}_{nm}^{(\nu)}+\left[\partial_{\nu}\Delta_{nm}\right]\left[\partial_{\gamma}\Delta_{nm}\right]\right]}{\left(\omega_1-\Delta_{nm}+i\Gamma\right)\left(\omega_2-\Delta_{nm}+i\Gamma\right)}-\frac{2\Delta_{nm}^{2}\mathcal{A}_{nm}^{(\mu)}\mathcal{A}_{mn}^{(\nu)}\mathcal{A}_{mn}^{(\gamma)}\mathcal{A}_{nm}^{(\delta)}}{\left(\omega_1-\Delta_{nm}+i\Gamma\right)\left(\omega_2-\Delta_{mn}+i\Gamma\right)} \nonumber\\[0.2cm]
 &-\frac{\mathcal{A}_{nm}^{(\delta)}\mathcal{A}_{mn}^{(\nu)}\left[\partial_{\mu}\Delta_{nm}\right]\left[\partial_{\gamma}\Delta_{nm}\right]}{\left(\omega_1-\Delta_{nm}+i\Gamma\right)\left(\omega_2+i\Gamma\right)} \text{ ,}
 \label{eq:Snm}
\end{align}

\begin{align}
\bar{T}_{nml,\mu\nu\gamma\delta}
=&\,
\frac{\mathcal{A}_{nm}^{(\delta)}}
{\omega_1-\Delta_{nm}+i\Gamma}
\Bigg\{
\frac{
\mathcal{A}_{ml}^{(\mu)}
\left[
\left(\partial_\nu\Delta_{lm}\right)\mathcal{A}_{ln}^{(\gamma)}
+
\left(\partial_\gamma\Delta_{nm}\right)\mathcal{A}_{ln}^{(\nu)}
\right]
}{
\omega_2-\Delta_{lm}+i\Gamma
}
\nonumber\\[0.2cm]
&+
\frac{
\mathcal{A}_{ln}^{(\mu)}
\left[
\left(\partial_\nu\Delta_{ln}\right)\mathcal{A}_{ml}^{(\gamma)}
+
\left(\partial_\gamma\Delta_{mn}\right)\mathcal{A}_{ml}^{(\nu)}
\right]
}{
\omega_2-\Delta_{nl}+i\Gamma
}
-
\frac{
\left(\partial_\mu\Delta_{lm}\right)
\mathcal{A}_{ml}^{(\nu)}
\mathcal{A}_{ln}^{(\gamma)}
+
\left(\partial_\mu\Delta_{ln}\right)
\mathcal{A}_{ml}^{(\gamma)}
\mathcal{A}_{ln}^{(\nu)}
}{
\omega_2+i\Gamma
}
\Bigg\}.
\label{eq:T}
\end{align}

\begin{align}
&\bar{U}_{nm,\mu\nu\gamma\delta}= \frac{\mathcal{A}_{mn}^{(\mu)}\mathcal{A}_{nm}^{(\delta)}\Delta_{nm}\left[s_{nn}^{(\nu\gamma)}-s_{mm}^{(\nu\gamma)}\right]}{\left(\omega_1-\Delta_{nm}+i\Gamma\right)\left(\omega_2-\Delta_{nm}+i\Gamma\right)}-\frac{i\left[\partial_{\mu}\Delta_{nm}\right]\mathcal{A}_{nm}^{(\delta)}s_{mn}^{(\nu\gamma)}}{\left(\omega_1-\Delta_{nm}+i\Gamma\right)\left(\omega_2+i\Gamma\right)}\text{ ,}
 \label{eq:Unm}
\end{align}

\begin{align}
    \bar{V}_{nml,\mu\nu\gamma\delta}&=\frac{\Delta_{ml}\mathcal{A}_{ml}^{(\mu)}\mathcal{A}_{nm}^{(\delta)}s_{ln}^{(\nu\gamma)}}{\left(\omega_1-\Delta_{nm}+i\Gamma\right)\left(\omega_2-\Delta_{lm}+i\Gamma\right)}-\frac{\Delta_{ln}\mathcal{A}_{ln}^{(\mu)}\mathcal{A}_{nm}^{(\delta)}s_{ml}^{(\nu\gamma)}}{\left(\omega_1-\Delta_{nm}+i\Gamma\right)\left(\omega_2-\Delta_{nl}+i\Gamma\right)} \text{ ,}
    \label{eq:V}
\end{align}

\begin{align}
    \bar{W}_{nmlp,\mu\nu\gamma\delta}&=\frac{\Delta_{ml}\Delta_{lp}\Delta_{pn}\mathcal{A}_{pn}^{(\gamma)}\mathcal{A}_{nm}^{(\delta)}}{\left(\omega_1-\Delta_{nm}+i\Gamma\right)}\left[\frac{\mathcal{A}_{ml}^{(\mu)}\mathcal{A}_{lp}^{(\nu)}}{\left(\omega_2-\Delta_{lm}+i\Gamma\right)}-\frac{\mathcal{A}_{lp}^{(\mu)}\mathcal{A}_{ml}^{(\nu)}}{\left(\omega_2-\Delta_{pl}+i\Gamma\right)}\right]\nonumber\\[0.2cm]
    &+\frac{\Delta_{ln}\Delta_{pl}\Delta_{mp}\mathcal{A}_{mp}^{(\gamma)}\mathcal{A}_{nm}^{(\delta)}}{\left(\omega_1-\Delta_{nm}+i\Gamma\right)}\left[\frac{\mathcal{A}_{ln}^{(\mu)}\mathcal{A}_{pl}^{(\nu)}}{\left(\omega_2-\Delta_{nl}+i\Gamma\right)}-\frac{\mathcal{A}_{pl}^{(\mu)}\mathcal{A}_{ln}^{(\nu)}}{\left(\omega_2-\Delta_{lp}+i\Gamma\right)}\right]+\left( l\leftrightarrow p \right) \text{ .}
\end{align}
\end{widetext}

\noindent Here, the $\mathbf{k}$ dependence has been omitted for readability, and the expression for $s_{nm}^{(\mu\nu)}(\mathbf{k})$ in terms of the non-Abelian Berry connection is shown in the Appendix \ref{app:chi_J}. 

We first discuss the spectral features arising from the purely paramagnetic part of the ABB response. The tensor $\bar{S}_{nm}$ gives rise to resonant features in both the first ($\omega_1>0 $ and $\omega_2\geq0$) and fourth ($\omega_1>0$ and $\omega_2<0$) quadrants of the frequency plane, with sharper peaks along the main diagonal ($\omega_1=\omega_2$; first term on the right-hand side of Eq.(\ref{eq:Snm})), the anti-diagonal ($\omega_2=-\omega_1$; second term) and the horizontal axis ($\omega_2=0$; third term). These different regions of the frequency plane probe distinct quantum-geometric quantities. More specifically, the features along the horizontal axis probe Cartesian components of the quadratic velocity-weighted pair-resolved quantum geometric tensor $\mathcal{Q}_{\nu\delta}^{m\rightarrow n}\left[\partial_{\mu}\Delta_{nm}\right]\left[\partial_{\gamma}\Delta_{nm}\right]$, while the anti-diagonal features probe products of four Berry connection matrix elements, $\mathcal{A}_{nm}^{(\mu)}\mathcal{A}_{mn}^{(\nu)}\mathcal{A}_{mn}^{(\gamma)}\mathcal{A}_{nm}^{(\delta)}$. In a two-band system, the latter can be expressed as bilinear combinations of components of the quantum geometric tensor of the valence band . Along the main diagonal, both quantities contribute. 

The tensors $\bar{T}_{nml}$ and $\bar{W}_{nmlp}$, by contrast, vanish identically in the two-band limit and therefore encode genuinely multiband contributions. Their energy denominators generate peaks that are not restricted to the main diagonal or anti-diagonal. Regarding their geometrical interpretation, $\bar{W}_{nmlp}$ also probes products of four components of the non-Abelian Berry connection. The tensor $\bar{T}_{nml}$ instead contains relative-velocity-weighted trilinear product interband Berry connections, such as  $\left(\partial_\nu\Delta_{lm}\right)\mathcal{A}_{ml}^{(\mu)}(\mathbf{k})\mathcal{A}_{ln}^{(\gamma)}(\mathbf{k})\mathcal{A}_{nm}^{(\delta)}(\mathbf{k})$. 

We now turn to the spectral features arising from the contact term $\chi_{T}^{\text{(ABB)}}(\omega_1,\omega_2)$. As can be seen from $\bar{U}_{nm}$ and $\bar{V}_{nml}$,  this contribution has no energy denominators that enforce features along $\omega_2=-\omega_1$ responsible for the intrinsic anti-diagonal echo feature of the paramagnetic response. The two-band contribution $\bar{U}_{nm}$ produces resonances along the main diagonal and horizontal axis on the frequency plane, whereas $\bar{V}_{nml}$, which is present only in multiband systems, gives rise to additional interband peaks. The corresponding matrix elements involve interband Berry connections and second-order current vertices $s_{nm}^{(\nu\gamma)}$, which can in turn be expressed in terms of band-energy derivatives and geometric quantities constructed from Berry connections and their covariant derivatives. For the idealized delta-function pulses considered here, the relative weight of $\chi_{\text{S}}^{(\text{ABB})}$ and $\chi_{\text{T}}^{(\text{ABB})}$ is not fixed without specifying a finite temporal pulse profile (see Appendix~\ref{app:chi_J}). We therefore analyze the spectral features of the two contributions separately.

The expressions for the AAB response $\chi^{(\text{AAB})}(\omega_1,\omega_2)$ and, more generically, for the 3DCS $\chi^{(3)}(\omega_1,\omega_2,\omega_3)$ are given in the Appendix \ref{app:chi_J}. Next, we particularize the expressions of the multidimensional optical susceptibilities for the two-band limit.

%-----
\subsection{Two-band systems}\label{subsec:two_bands}

The optical responses simplify considerably in the two-band limit. As we show below, for a two-band insulator, the pair-resolved quantum geometric tensor $\mathcal{Q}_{\mu\nu}^{m\rightarrow n}(\mathbf{k})$, with $m$ and $n$ denoting valence and conduction bands, respectively, coincides with the quantum geometric tensor of the valence band. As a consequence, the geometric content of the optical responses can be directly interpreted in terms of the valence-band quantum metric and Berry curvature. 

The quantum geometric tensor of an isolated band is defined as~\cite{provost1980riemannian,shapere1989geometric}
\begin{align}
    \mathcal{Q}_{\mu\nu}^{(n)}(\mathbf{k})&=\left\langle\left.\partial_{\mu}u_n\right|\partial_{\nu}u_n\right\rangle-\left\langle\left.\partial_{\mu}u_n\right|u_n\right\rangle\left\langle\left.u_n\right|\partial_{\nu}u_n\right\rangle\nonumber\\[0.2cm]
    &=\sum\limits_{m\neq n}\mathcal{A}_{nm}^{(\mu)}(\mathbf{k})\mathcal{A}_{mn}^{(\nu)}(\mathbf{k})=g_{\mu\nu}^{(n)}(\mathbf{k})-\frac{i}{2}\Omega_{\mu\nu}^{(n)}(\mathbf{k}) \text{ ,}
    \label{eq:QGT}
\end{align}

\noindent where the $\mathbf{k}$ dependence of the cell-periodic Bloch wave functions was omitted to shorten the notation. In Eq.(\ref{eq:QGT}), $g_{\mu\nu}^{(n)}(\mathbf{k})$ is a real symmetric tensor that encodes information about the infinitesimal distance between the cell-periodic Bloch states, and is called quantum metric. Additionally, $\Omega_{\mu\nu}^{(n)}(\mathbf{k})$ is the Berry curvature of band $n$. For a 2D insulator, the integral of $\Omega_{\mu\nu}^{(n)}(\mathbf{k})$ in the Brillouin yield the Chern number, which characterizes the band topology of the system. A system with finite $g_{\mu\nu}^{(n)}(\mathbf{k})$ is said to have a non-trivial quantum geometry even if it has trivial topology~\cite{torma2023essay}.  

For a two-band system, the sum over $m\neq n$ in Eq.(\ref{eq:QGT}) simplifies to $\mathcal{Q}^{(1)}_{\mu\nu}(\mathbf{k})=\mathcal{A}_{12}^{(\mu)}(\mathbf{k})\mathcal{A}_{21}^{(\nu)}(\mathbf{k})$, which is equivalent to $\mathcal{Q}_{\mu\nu}^{1\rightarrow 2}(\mathbf{k})$. We can thus rewrite Eq.(\ref{eq:chi_1_omega_A}) as 
\begin{equation}
    \chi_{\mu\nu}^{(1)}(\omega>0)=-\frac{1}{V}\sum\limits_{\mathbf{k}}\frac{f_{12}(\mathbf{k})\Delta_{21}^2(\mathbf{k})\mathcal{Q}_{\mu\nu}^{(1)}(\mathbf{k})}{\omega-\Delta_{21}(\mathbf{k})+i\Gamma} \text{ ,}
    \label{eq:chi_1_Q}
\end{equation}

\noindent with $\chi_{\mu\nu}^{(1)}(-\omega)=\left[\chi_{\mu\nu}^{(1)}(\omega)\right]^*$, and $\omega>0$. Note that the quantum geometric tensors of bands $1$ and $2$ are related by $\mathcal{Q}_{\mu\nu}^{(2)}(\mathbf{k})=\mathcal{A}_{21}^{(\mu)}(\mathbf{k})\mathcal{A}_{12}^{(\nu)}(\mathbf{k})=\left[\mathcal{Q}_{\mu\nu}^{(1)}(\mathbf{k})\right]^*$. Therefore, $g_{\mu\nu}^{(2)}(\mathbf{k})=g_{\mu\nu}^{(1)}(\mathbf{k})$ and $\Omega_{\mu\nu}^{(2)}(\mathbf{k})=-\Omega_{\mu\nu}^{(1)}(\mathbf{k})$.

In the case of the multidimensional susceptibilities, only the terms involving $\bar{P}_{nm}(\mathbf{k})$, $\bar{S}_{nm}(\mathbf{k})$ and $\bar{U}_{nm}$ yield finite contributions to $\chi^{(2)}(\omega_1,\omega_2)$ [Eq.(\ref{eq:chi_2main})], $\chi^{(\text{ABB})}_S(\omega_1,\omega_2)$ [Eq.(\ref{eq:chi_ABB})] and $\chi^{(\text{ABB})}_T(\omega_1,\omega_2)$ [Eq.(\ref{eq:chi_ABBT})]. These tensors take the following simpler expressions:
\begin{align}
     \bar{P}_{21,\mu\nu\gamma}&=\frac{\left[\partial_\nu\Delta_{21}\right]\mathcal{Q}_{\mu\gamma}^{(1)}}{\left(\omega_1-\Delta_{21}+i\Gamma\right)\left(\omega_2-\Delta_{21}+i\Gamma\right)}\nonumber\\[0.2cm]
     &-\frac{\left[\partial_\mu\Delta_{21}\right]\mathcal{Q}_{\nu\gamma}^{(1)}}{\left(\omega_1-\Delta_{21}+i\Gamma\right)\left(\omega_2+i\Gamma\right)}\text{ ,}
     \label{eq:P_2bands}
 \end{align}

\begin{align}
 &\bar{S}_{21,\mu\nu\gamma\delta}= -\frac{2\Delta_{21}^{2}\mathcal{Q}^{(1)}_{\nu\mu}\mathcal{Q}^{(1)}_{\gamma\delta}}{\left(\omega_1-\Delta_{21}+i\Gamma\right)\left(\omega_2+\Delta_{21}+i\Gamma\right)}\nonumber\\[0.2cm]
 &-\frac{\mathcal{Q}^{(1)}_{\nu\delta}\left[\partial_{\mu}\Delta_{21}\right]\left[\partial_{\gamma}\Delta_{21}\right]}{\left(\omega_1-\Delta_{21}+i\Gamma\right)\left(\omega_2+i\Gamma\right)}\nonumber\\[0,2cm]
 &+\frac{\mathcal{Q}^{(1)}_{\mu\delta}\left[2\Delta_{21}^2\mathcal{Q}^{(1)}_{\gamma\nu}+\left(\partial_{\nu}\Delta_{21}\right)\left(\partial_{\gamma}\Delta_{21}\right)\right]}{\left(\omega_1-\Delta_{21}+i\Gamma\right)\left(\omega_2-\Delta_{21}+i\Gamma\right)}\text{ ,}
 \label{eq:S_2bands}
 \end{align}

\noindent and 
 \begin{align}
     &\bar{U}_{21,\mu\nu\gamma\delta}=\frac{-i\left[\partial_{\mu}\Delta_{21}\right]\mathcal{A}_{21}^{(\delta)}s_{12}^{(\nu\gamma)}}{\left(\omega_1-\Delta_{21}+i\Gamma\right)\left(\omega_2+i\Gamma\right)}\nonumber\\[0.2cm]
 &+\frac{4\mathcal{Q}^{(1)}_{\mu\delta}\Delta_{21}\left[-\Delta_{21}g_{\nu\gamma}^{(1)}+\frac{1}{4}\partial_{\nu}\partial_{\gamma}\Delta_{21}\right]}{\left(\omega_1-\Delta_{21}+i\Gamma\right)\left(\omega_2-\Delta_{21}+i\Gamma\right)}\text{ ,}
 \label{eq:U_2bands}
 \end{align}

\noindent where the $\mathbf{k}$-dependence was omitted to shorten the notation.

\begin{figure*}[t!]
\includegraphics[width=0.99\linewidth]{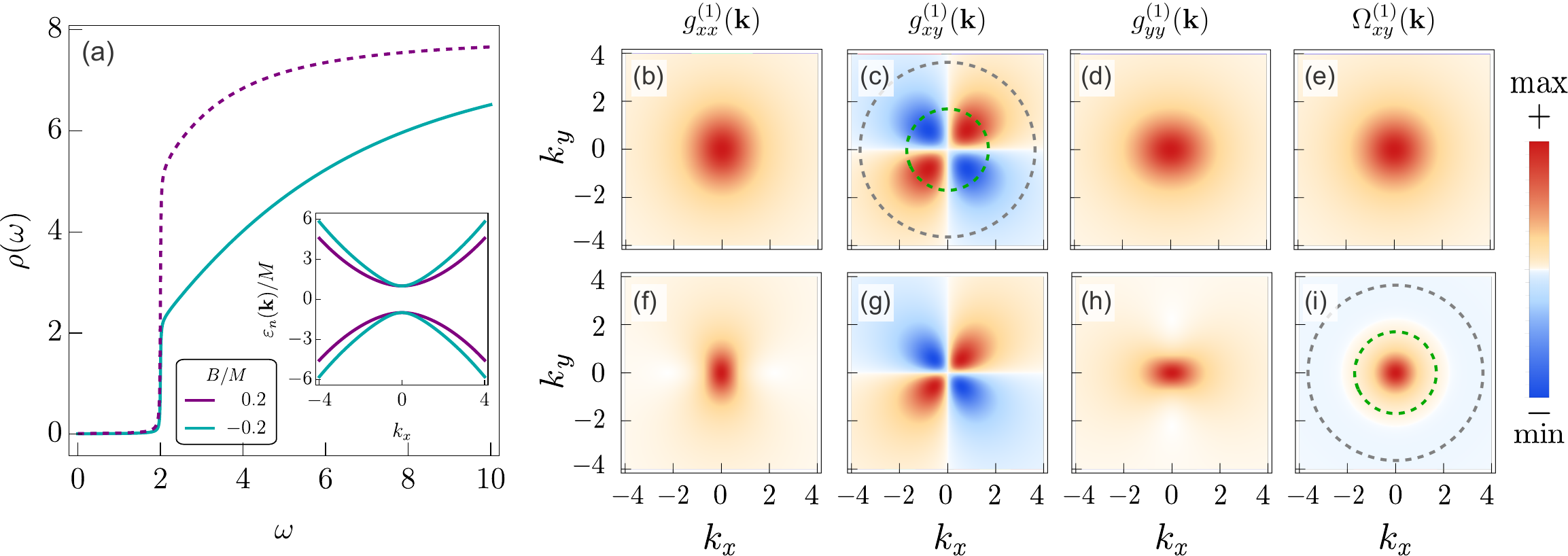}
\caption{Band structure and band geometry of Dirac fermions. (a) Joint density of states and band dispersion (inset) in the topologically trivial ($B/M=-0.2$) and non-trivial ($B/M=0.2$) regimes. Panels (b)-(i) show the profile of the components of the quantum metric and Berry curvature, normalized to their maximum absolute values, for $B/M=0.2$ (top row) and $B/M=-0.2$ (bottom row). Additionally, panels (c) and (i) show the optically active region for $\omega=3.5$ (dashed green curve) and $\omega=8$ (dashed gray curve).}
\label{fig:band_untilted}
\end{figure*}

Although the concept of quantum geometric tensor can be extended to a manifold of occupied bands~\cite{ahn2022riemannian,ahn2020low}, the energy denominator of the optical susceptibilities in this case can no longer be separated from the products of Berry connection matrix elements entering  the numerators. Therefore, a direct connection with the quantum geometric tensor of the occupied manifold is not exact in a multiband system. However, Eqs.(\ref{eq:chi_1_Q})-(\ref{eq:S_2bands}) are expected to be good approximations for the low-frequency regime response at frequencies near band gaps. 

%----

\section{Results}\label{sec:results}

In this section, we present the theoretically calculated multidimensional optical susceptibilities for two prototypical Chern-insulator models. We first analyze a two-dimensional continuum Dirac model, for which both the quantum geometric tensor and the optical response can be obtained analytically. This provides a simple setting to the quantum-geometric interpretation of the features in the 2DCS and also serves as a useful building block for the responses of more complex insulators. We then consider a square-lattice Chern insulator, where the more intricate band dispersion leads to richer spectroscopic features. In the results shown next, we set $\Gamma=0.01$. 

\subsection{Continuous Dirac model}\label{sec:Dirac}

A minimal model in two dimensions that exhibits non-trivial topology consists of massive Dirac fermions in continuous space. The corresponding first-quantized Hamiltonian is~\cite{BernevigHughes2013,franz2013topological}
\begin{equation}
\hat{h}\left(\mathbf{k}\right)=v_x k_x\sigma_x+v_y k_y\sigma_y+\mathcal{M}(\mathbf{k})\sigma_z -\mu\sigma_0\text{ ,}
\label{eq:h_Dirac}
\end{equation}

\begin{figure}[b!]
    \centering
\includegraphics[width=0.99\linewidth]{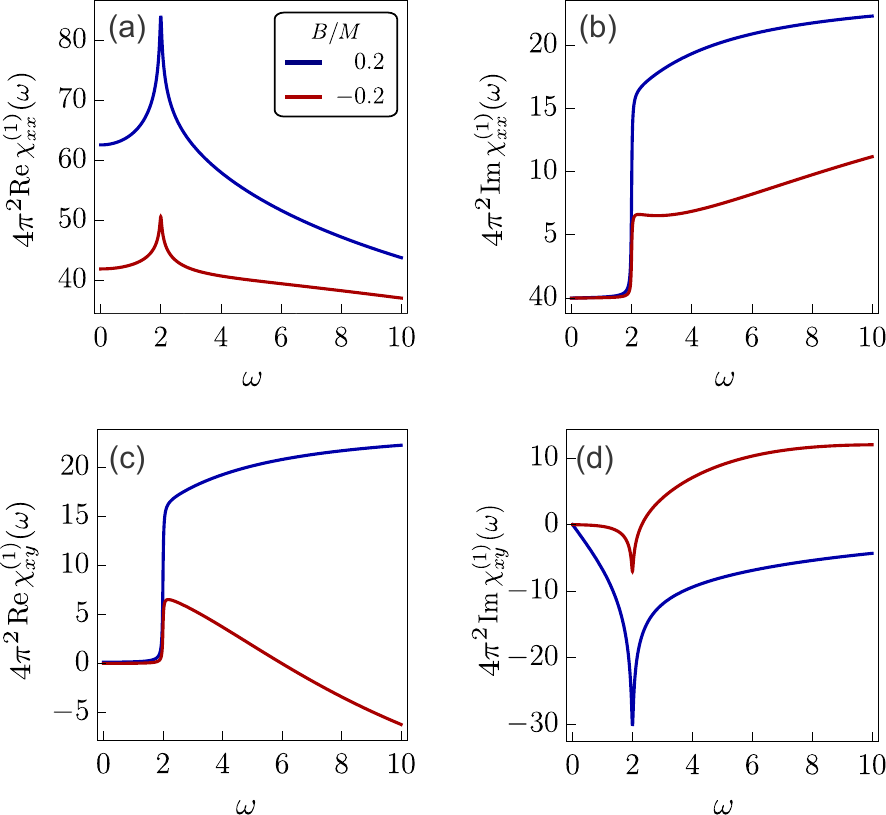}
    \caption{Linear response of the continuous Dirac fermion model. The panels show the real and imaginary parts of the two distinct components of $\chi^{(1)}(\omega)$ for the model defined in Eq.(\ref{eq:h_Dirac}), contrasting the features in the topologically trivial ($B/M=-0.2$; red curve) and non-trivial ($B/M=0.2$; blue curve) regimes. Qualitative differences between the two regimes are manifested in the $\chi_{xy}^{(1)}(\omega)$ component.}
    \label{fig:Linear_untilted}
\end{figure}

\noindent where $v_x$ and $v_y$ are the Cartesian components of the Dirac velocity, and the chemical potential  $\mu$ controls the occupation of the bands. Additionally, $\sigma_j$ (with $j=0,x,y,z$) are the Pauli matrices in the two-dimensional state space of this model, which may correspond, for instance, to spin-1/2 or opposite-parity orbitals, depending on physical realization.

We focus on charge neutrality ($\mu=0$), where the system is insulating with a minimum band gap at $\mathbf{k}=0$, as shown in the inset of Fig.\ref{fig:band_untilted}(a). We further set $v_x=v_y\equiv v$, resulting in an isotropic dispersion, and adopt units such that $\hbar=a=1$, where $a$ sets the characteristic length scale. With this convention, $v$ has dimensions of energy and $\mathbf{k}$ is dimensionless. Throughout, all quantities with dimensions of energy are expressed in units of $v$. The momentum-dependent mass term of this model $\mathcal{M}(\mathbf{k})=M-Bk^2$, with $k=\sqrt{k_x^2+k_y^2}$, contains the Dirac $M$ mass and the Newtonian mass $B$~\cite{franz2013topological}. The relative sign of these parameters determines the band topology: for $B/M>0$, the system is in the topologically nontrivial regime, with valence-band Chern number $C=1$, whereas for $B/M<0$ it is a trivial insulator with $C=0$~\cite{BernevigHughes2013,franz2013topological}. Throughout, we set $M=1$. We next show that the linear optical susceptibility is primarily sensitive to the Berry curvature integrated over the optically active region, whereas the 2DCS response resolves sign changes in bilinear combinations of the quantum geometric tensor. 

We focus first on the paramagnetic part of the linear response tensor. As discussed in Sec.\ref{sec:MDCS_theory}, the contact term contains a generically finite, frequency-independent interband Fermi-sea contribution in an insulator. We subtract this constant contribution from the responses shown below in order to emphasize their frequency-dependent resonant features. The paramagnetic part of the linear response tensor has two distinct components within this model. The isotropy of the energy dispersion gives $\chi_{xx}^{(1)}(\omega)=\chi_{yy}^{(1)}(\omega)$. Moreover, time-reversal symmetry (TRS) is broken, allowing for a finite Hall response, $\chi_{xy}^{(1)}(\omega)=-\chi_{yx}^{(1)}(\omega)$. Note that, if the internal degrees of freedom correspond to spin-1/2, the time-reversal operator is represented by $\mathcal{T}=i\sigma_y\mathcal{K}$, where $\mathcal{K}$ denotes complex conjugation, and TRS is broken by a finite $M$ or $B$. In a purely orbital model, on the other hand, $\mathcal{T}=\sigma_0\mathcal{K}$ and TRS is broken by the term proportional to $k_x\sigma_x$ in the Hamiltonian~\cite{BernevigHughes2013}.  

The two independent components of the linear response tensor are shown in Fig. \ref{fig:Linear_untilted} for two values of $B/M$ representative of the topologically trivial ($B/M=-0.2$) and nontrivial ($B/M=0.2$) regimes. Sharp features, including peaks and jumps, appear at $\omega=2$, corresponding to the onset of interband transitions at the minimum band gap $\Delta_{21}^{(\text{min})}=2M$ [see Fig.\ref{fig:band_untilted}(a)], and to the associated enhancement of the joint density of states (JDOS). However, the JDOS alone does not determine all spectroscopic features of the linear response. As shown in Sec.\ref{subsec:two_bands}, $\chi_{\mu\nu}^{(1)}(\omega)$ probes $\Delta_{21}^{2}(\mathbf{k})\mathcal{Q}_{\mu\nu}^{(1)}(\mathbf{k})$ integrated over the optically active region (OAR).  Owing to the rotational symmetry of the present model, the corresponding resonant constant-energy contours are concentric circles centered at $\mathbf{k}=0$, whose radii are determined by the resonance condition $\Delta{21}(k)=\omega$ (see Fig.\ref{fig:band_untilted}). We recall that $\mathcal{Q}_{\mu\nu}^{(1)}(\mathbf{k})=g_{\mu\nu}^{(1)}(\mathbf{k})-\frac{i}{2}\Omega_{\mu\nu}^{(1)}(\mathbf{k})$ is the quantum geometric tensor of the valence band, which assumes a simple analytic expression for this model,
\begin{align}
   & g_{\mu\nu}^{(1)}(\mathbf{k})=\frac{\left(4MB-1\right)k_\mu k_\nu+f(k)\delta_{\mu,\nu}}{4f^2(k)}\text{ ,}\label{eq:g_2D}\\[0,2cm]
    & \Omega_{xy}^{(1)}(\mathbf{k})=\frac{\left[Bk^2+M\right]}{2f^{\frac{3}{2}}(k)} \text{ ,}
    \label{eq:Omega_2D}
\end{align}

\noindent with $f(k)=k^2+\left(M-Bk^2\right)^2$,  $\Omega_{yx}^{(1)}(\mathbf{k})=-\Omega_{xy}^{(1)}(\mathbf{k})$, and the diagonal components of the Berry curvature tensor vanish identically. Their momentum-space profiles are shown in Fig.\ref{fig:band_untilted}(b)-(i) for the topologically non-trivial ($B/M=0.2$; first row) and trivial ($B/M=-0.2$; second row) regimes. 

\begin{figure}[t!]
    \centering
    \includegraphics[width=0.99\linewidth]{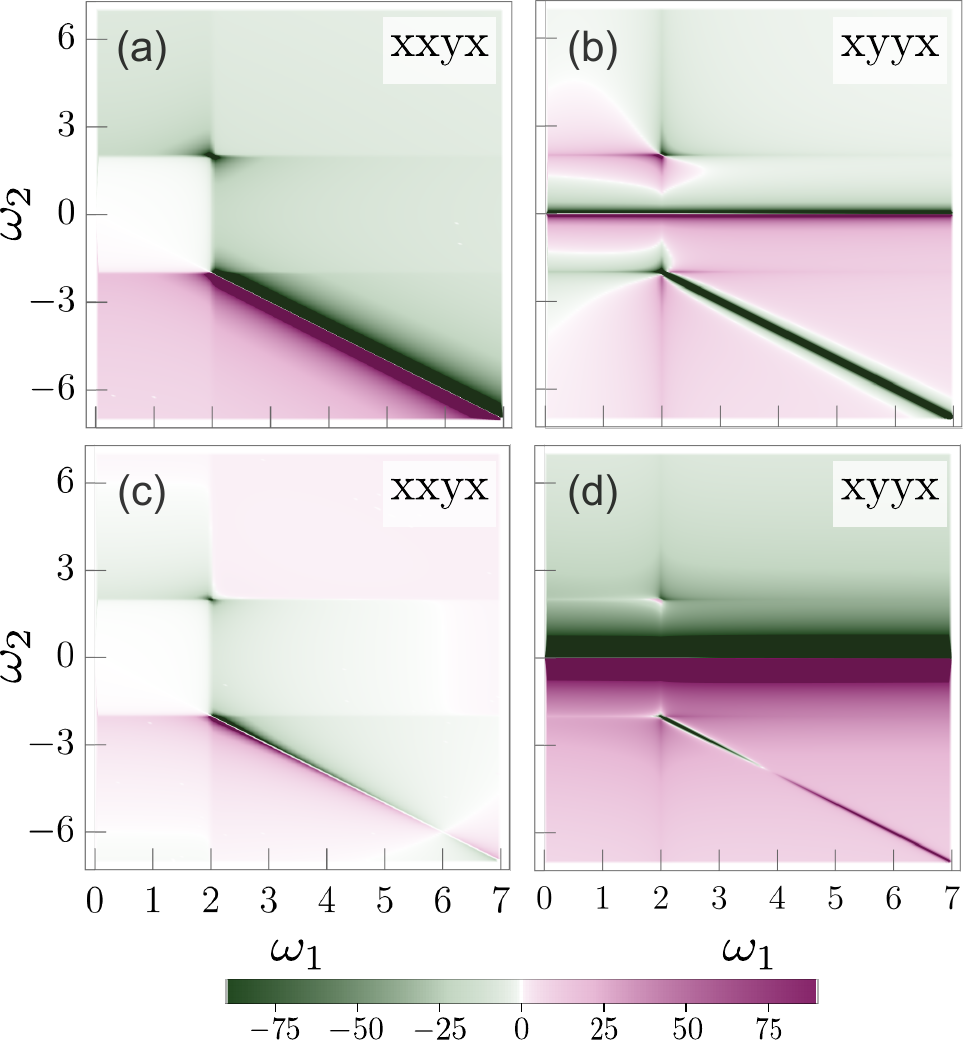}
    \caption{Imaginary part of $4\pi^2\chi_{xxyx}^{\text{(ABB)}}(\omega_1,\omega_2)$ (left column) and $4\pi^2\chi_{xyyx}^{\text{(ABB)}}(\omega_1,\omega_2)$ (right column). The responses shown in (a) and (b) were calculated in the topologically non-trivial regime with $B/M=0.2$, while  (c) and (d) were obtained in the topologically trivial regime with $B/M=-0.2$.}
    \label{fig:Dirac_2DCS}
\end{figure}

The quantum metric $g_{xx}^{(1)}(\mathbf{k})$ has qualitatively similar momentum-space profiles for $B/M<0$ and $B/M>0$. As a consequence, the differences between topological and trivial regimes in $\chi_{xx}^{(1)}(\omega)$ are mainly quantitative. The scenario is different for $\chi_{xy}^{(1)}(\omega)$, which shows two distinctive signatures of the geometry and topology of the electronic states. First, the real part of $\chi_{xy}^{(1)}(\omega)$ changes sign as a function of $\omega$ in the trivial regime, but not in the topological regime [Fig.~\ref{fig:Linear_untilted}(c)]. This follows from the behavior of the Berry curvature of this system. As shown in Fig.\ref{fig:band_untilted}(e), $\Omega_{xy}^{(1)}(\omega)$ has a monopole-like structure in the topological regime, being peaked at the position of the minimum band gap ($\mathbf{k}=0$). As a consequence, its contribution integrated over the OAR, retains the same sign as the resonant energy contour moves towards larger momenta. In the trivial regime, on the other hand, $\Omega_{xy}^{(1)}(\omega)$ is still large and positive at $\mathbf{k}=0$, but changes sign for larger momenta [see Fig.\ref{fig:band_untilted}(i)]. Therefore, as $\omega$ increases, the OAR crosses from the positive to the negative region of $\Omega_{xy}^{(1)}(\mathbf{k})$, resulting in the sign change in $\text{Re}\chi_{xy}^{(1)}(\omega)$.

A second signature appears in the low-frequency behavior of $\text{Im}\chi_{xy}^{(1)}(\omega)$. From Eq.(\ref{eq:chi_1_Q}), the leading low-frequency contribution is proportional to the Chern number, $\lim\limits_{\omega\rightarrow 0^+}\text{Im}\chi_{xy}^{(1)}(\omega)=-\frac{\omega C}{2\pi}$. Thus the linear-in-$\omega$ contribution vanishes in the trivial regime, whereas in the topological regime $\text{Im}\chi_{xy}^{(1)}(\omega)$ approaches zero linearly with a slope determined by the Chern number. 

We now turn to the 2DCS of this model. The lowest order 2DCS, $\chi_{\mu\nu\gamma}^{(2)}(\omega_1,\omega_2)$ vanishes identically due to spatial inversion symmetry of the model. We therefore consider the third-order response and focus on the ABB 2DCS. In the following, we discuss primarily the spectral features of the paramagnetic contribution defined in Eq.(\ref{eq:chi_ABB}). The contact contribution is concentrated mainly along the main diagonal and the horizontal axis, where its resonances occur, and is discussed separately in Appendix~\ref{app:contact_Dirac}.

The symmetries of the model impose constraints on the components of the third-order tensor. Namely,  $xxxx=yyyy$, $xxxy=-yyyx$, $xxyx=-yyxy$, $xyxx=-yxyy$, $yxxx=-xyyy$, $xxyy=yyxx$, $xyxy=yxyx$, and $xyyx=yxxy$, where we used a shorthand notation in which only the Cartesian indices are displayed. The $xxyx$ and $xyyx$ components are shown in Fig.\ref{fig:Dirac_2DCS}. Both exhibit clear qualitative differences between the topologically non-trivial (upper row) and trivial (lower row) regimes, particularly along the anti-diagonal $\omega_2=-\omega_1$. As shown below, these differences can be directly related to the quantum geometry of the Bloch states. 

\begin{figure*}[t!]
    \centering
\includegraphics[width=0.75\linewidth]{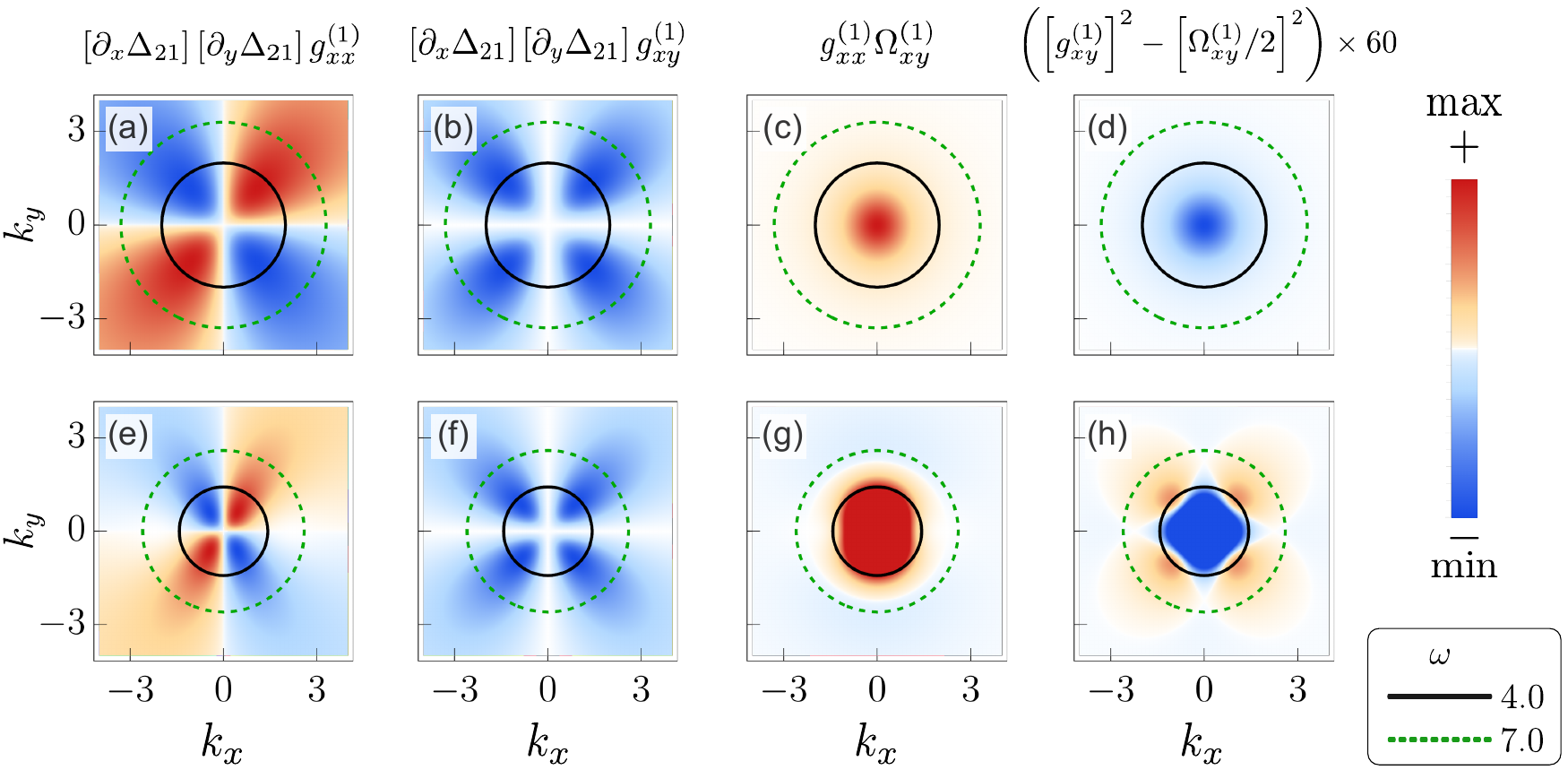}
    \caption{Momentum space profile of the quantum geometric quantities manifested in the components of the ABB 2DCS of Fig.\ref{fig:Dirac_2DCS}.}
    \label{fig:Dirac_geometry_part2}
\end{figure*}

As discussed in Sec.\ref{sec:connection_geometry}, distinct regions of the $\omega_1\times \omega_2$ plane probe different quantum geometric quantities. In particular, the horizontal axis ($\omega_2=0$) of the $xxyx$ response probes the quadratic relative-velocity weighted quantum metric $\left[\partial_{x}\Delta_{21}(\mathbf{k})\right]\left[\partial_{y}\Delta_{21}(\mathbf{k})\right]g_{xx}^{(1)}(\mathbf{k})$, whose momentum space profile is shown in Fig.\ref{fig:Dirac_geometry_part2}(a) and  Fig.\ref{fig:Dirac_geometry_part2}(e) for the topologically non-trivial and trivial regimes, respectively. In both cases, this quantity integrates to zero over the OAR as a consequence of its angular structure, resulting in the absence of a resonant feature along the horizontal axis of this component. Additionally, the minimum band gap of the model is clearly resolved in the $xxyx$ response, with no resonant spectral weight in the region $\omega_1,\left|\omega_2\right|<2M$. 

The $xyyx$ component, on the other hand, has a finite signal along the horizontal axis, which probes $\left[\partial_{x}\Delta_{21}(\mathbf{k})\right]\left[\partial_{y}\Delta_{21}(\mathbf{k})\right]\left[\mathcal{Q}_{xy}^{(1)}(\mathbf{k})\right]^*$. Only the real part of this geometric quantity survives integration over the OAR, since $\left[\partial_{x}\Delta_{21}(\mathbf{k})\right]\left[\partial_{y}\Delta_{21}(\mathbf{k})\right]\Omega_{xy}^{(1)}(\mathbf{k})$ integrates to zero by symmetry. The momentum space profile of $\left[\partial_{x}\Delta_{21}(\mathbf{k})\right]\left[\partial_{y}\Delta_{21}(\mathbf{k})\right]g_{xy}^{(1)}(\mathbf{k})$ is shown in Fig.\ref{fig:Dirac_geometry_part2}(b) and Fig.\ref{fig:Dirac_geometry_part2}(f) for the topologically non-trivial and trivial regimes, respectively. Their qualitative similarity in the two regimes results in correspondingly similar spectral features along the horizontal axis of the $xyyx$ response. 

The distinction between the two topological regimes is much more pronounced along the anti-diagonal $\omega_2=-\omega_1$. For the $xxyx$ component, this echo signal probes the bilinear combination $\mathcal{Q}_{xx}^{(1)}(\mathbf{k})\mathcal{Q}_{yx}^{(1)}(\mathbf{k})=g_{xx}^{(1)}(\mathbf{k})\left[\mathcal{Q}_{xy}^{(1)}(\mathbf{k})\right]^*$. Its imaginary part is proportional to $g_{xx}^{(1)}(\mathbf{k})\Omega_{xy}^{(1)}(\mathbf{k})$, whose momentum-space profile is shown in Fig.\ref{fig:Dirac_geometry_part2}(c) and Fig.\ref{fig:Dirac_geometry_part2}(g), and provides the dominant geometric contribution to the anti-diagonal sector. Since $g_{xx}^{(1)}(\mathbf{k})$ is always positive, the sign structure of this quantity follows that of the Berry curvature. It, therefore, changes sign at finite momentum in the trivial regime, but not in the topological regime. As a result, the echo signal changes sign in the trivial regime as the resonant contour moves towards larger momenta with increasing $\omega_1$. 

A similar sign change occurs in the  $xyyx$ echo in the trivial regime. In this case, the relevant quantum geometric quantity is $\mathcal{Q}_{yx}^{(1)}(\mathbf{k})\mathcal{Q}_{yx}^{(1)}(\mathbf{k})$, whose real part is $\left[g_{xy}^{(1)}(\mathbf{k})\right]^2-\left[\Omega_{xy}^{(1)}(\mathbf{k})/2\right]^2$. The OAR-integrated value of this quantity changes sign in the trivial regime [compare Fig.\ref{fig:Dirac_geometry_part2}(d) and \ref{fig:Dirac_geometry_part2}(h)], accounting for the corresponding sign reversal of the echo. The imaginary part, $\text{Im}\left[\mathcal{Q}_{yx}^{(1)}(\mathbf{k})\mathcal{Q}_{yx}^{(1)}(\mathbf{k})\right]=g_{xy}^{(1)}(\mathbf{k})\Omega_{xy}^{(1)}(\mathbf{k})$ vanishes upon integration over the OAR by symmetry. The same qualitative echo sign reversal is found in the imaginary parts of both $xxyx$ and $xyyx$ responses in the trivial regime. 

Finally, because the continuous Dirac model has an unbounded dispersion, we introduce a momentum cutoff $\Lambda$ in evaluating the momentum-space integrals. For all the results presented in this subsection, we used $\Lambda =20$, so that $\Delta_{21}(\Lambda)\gg\omega_{\text{max}}$, where $\omega_{max}=10$ the largest frequency  considered in our calculations. Different choices of $\Lambda$ change quantitative aspects of the responses, but leaves the qualitative spectral features discussed above unchanged.

\subsection{Topological insulator on a square lattice}\label{sec:results_square}

In this section, we present the linear and multidimensional optical susceptibilities of a two-dimensional Chern insulator on a square lattice. In contrast to the continuum Dirac model, the lattice regularization provides a compact Brillouin zone and eliminates the need for a momentum cutoff. The richer band dispersion and quantum geometry also give rise to additional spectroscopic features, which can be understood from the evolution of the optically active regions (OARs) across the Brillouin zone.

\begin{figure}[b!]
    \centering
\includegraphics[width=0.95\linewidth]{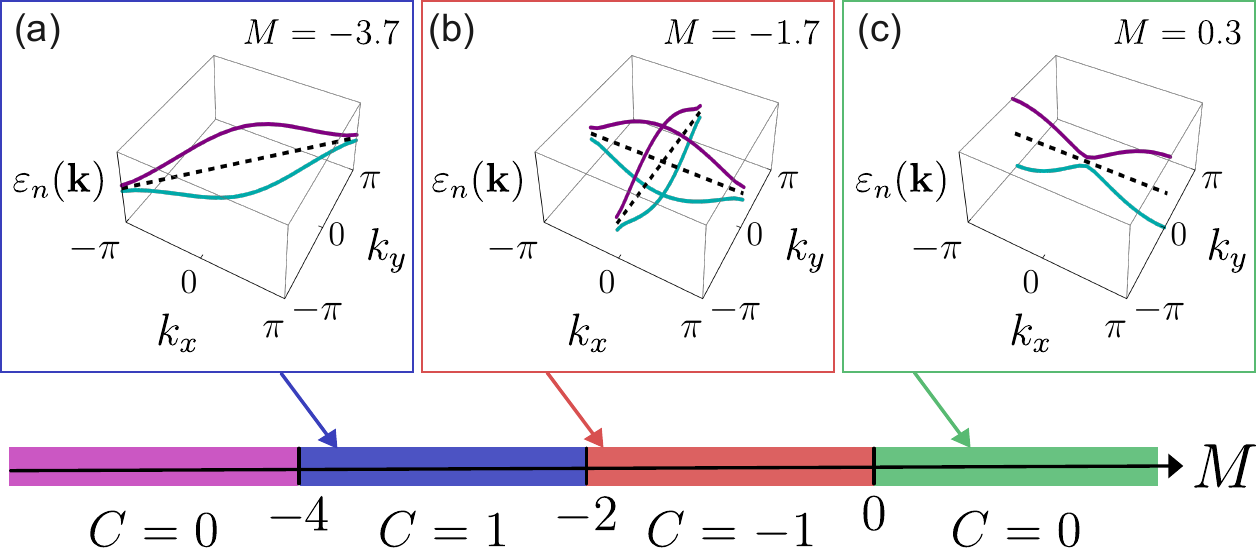}
    \caption{Band structure and topology of a Chern insulator on a square lattice. Panels (a)-(c) show cuts of representative band dispersions in the distinct topological regimes supported by the model. In (a), the system has Chern number $C=1$ with minimum band gap at $\mathbf{k}=\pm(\pi,\pi)$. In (b), $C=-1$ and the gap minima are located at the points $\mathbf{k}=(0,\pm\pi)$, $(\pm\pi,0)$. In (c), the system is topologically trivial ($C=0$), with the minimum band gap at the center of the Brillouin zone.}
 \label{fig:bands_square}
\end{figure}

The single-particle Hamiltonian is
\begin{equation}
    \hat{h}(\mathbf{k})=\sin k_x\sigma_x+\sin k_y\sigma_y+\xi(k_x,k_y)\sigma_z-\mu\sigma_0 \text{ ,}
    \label{eq:h_square}
\end{equation}

\noindent with
\begin{equation}
    \xi(k_x,k_y)=2+M-\cos k_x-\cos k_y \text{ .}
\end{equation}

\begin{figure}[t!]
    \centering
    \includegraphics[width=0.92\linewidth]{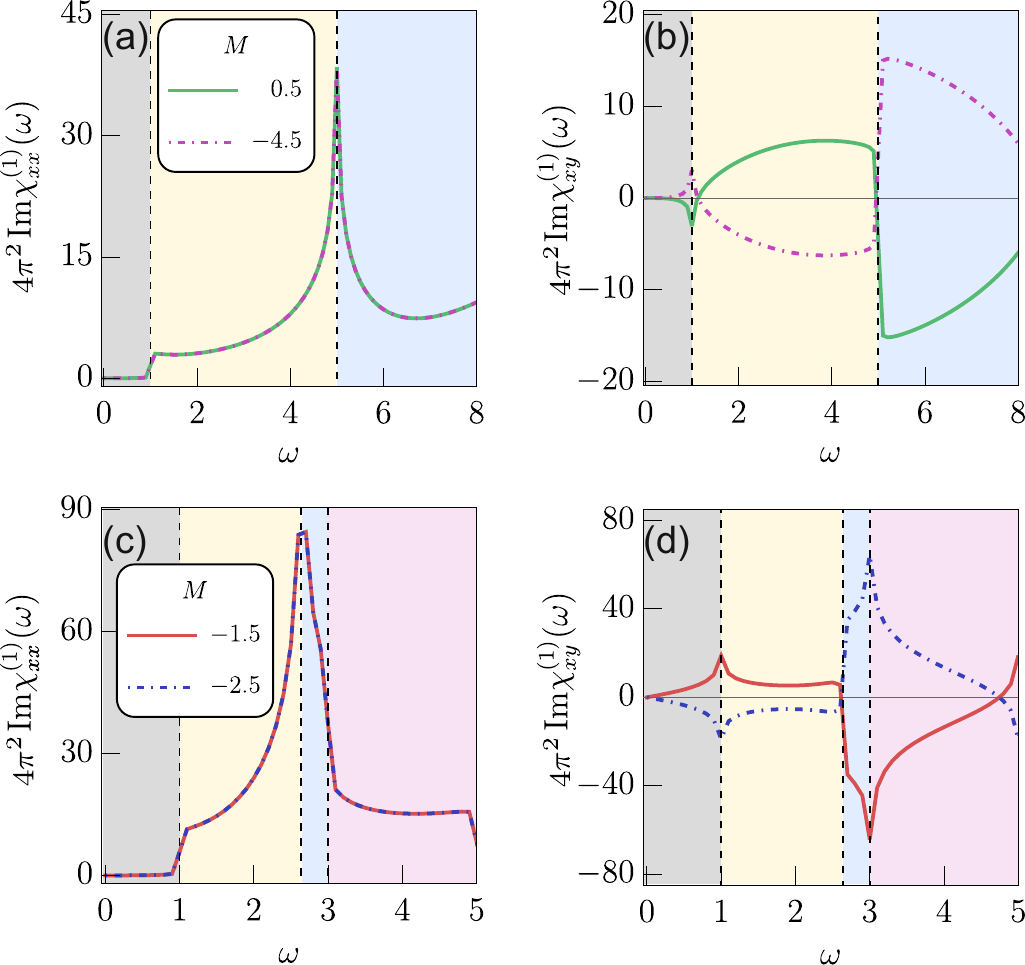}
    \caption{Linear optical response of a Chern insulator in a square lattice for values of $M$ representative of each region of the phase diagram shown in Fig.\ref{fig:bands_square}. The first row shows the imaginary part of (a) $\chi_{xx}^{(1)}(\omega)$ and (b) $\chi_{xy}^{(1)}(\omega)$ in the topologically trivial regime, while the second row shows the same components in the topologically nontrivial regimes. The shaded regions indicate different topologies of the optically active region as a function of $\omega$. We follow the same color code as in Fig.\ref{fig:OAR_square} shown in the Appendix \ref{app:JDOS_square}.}
\label{fig:linear_square}
\end{figure}

\noindent This model is a minimal lattice generalization of the continuum Dirac model considered in Sec.\ref{sec:Dirac}, and describes spinless electrons on a square lattice with two opposite-parity orbitals per site~\cite{BernevigHughes2013}.

The mass $M$ tunes band topology. The system is topologically trivial, with Chern number $C=0$ for $M>0$ and $M<-4$ and non-trivial otherwise, as illustrated in Fig.\ref{fig:bands_square}. The valence-band Chern number can assume the values $C=-1$ for $-2<M<0$, and  $C=1$ for $-4<M<-2$. At the critical masses $M_{c,1}=0$, $M_{c,2}=-2$, and $M_{c,3}=-4$, the electronic gap closes at different high-symmetry points of the Brillouin zone, as shown in Fig.\ref{fig:bands_square}. The quantum metric and the Berry curvature become strongly enhanced near these points and therefore play an important role in the optical responses close to the topological transitions.

The imaginary part of the two independent components of the linear optical susceptibility of this model are shown in Fig.\ref{fig:linear_square}. The frequency range in each panel is set by the bandwidth, which depends on $M$. The spectra display several sharp peaks and jumps associated with singularities of the JDOS. These occur at frequencies for which the topology of the OAR changes (indicated by the dashed lines in Fig.\ref{fig:linear_square}), in close analogy with Lifshitz transitions in metals (see Appendix \ref{app:JDOS_square}). 

\begin{figure*}[t!]
    \centering
    \includegraphics[width=0.7\linewidth]{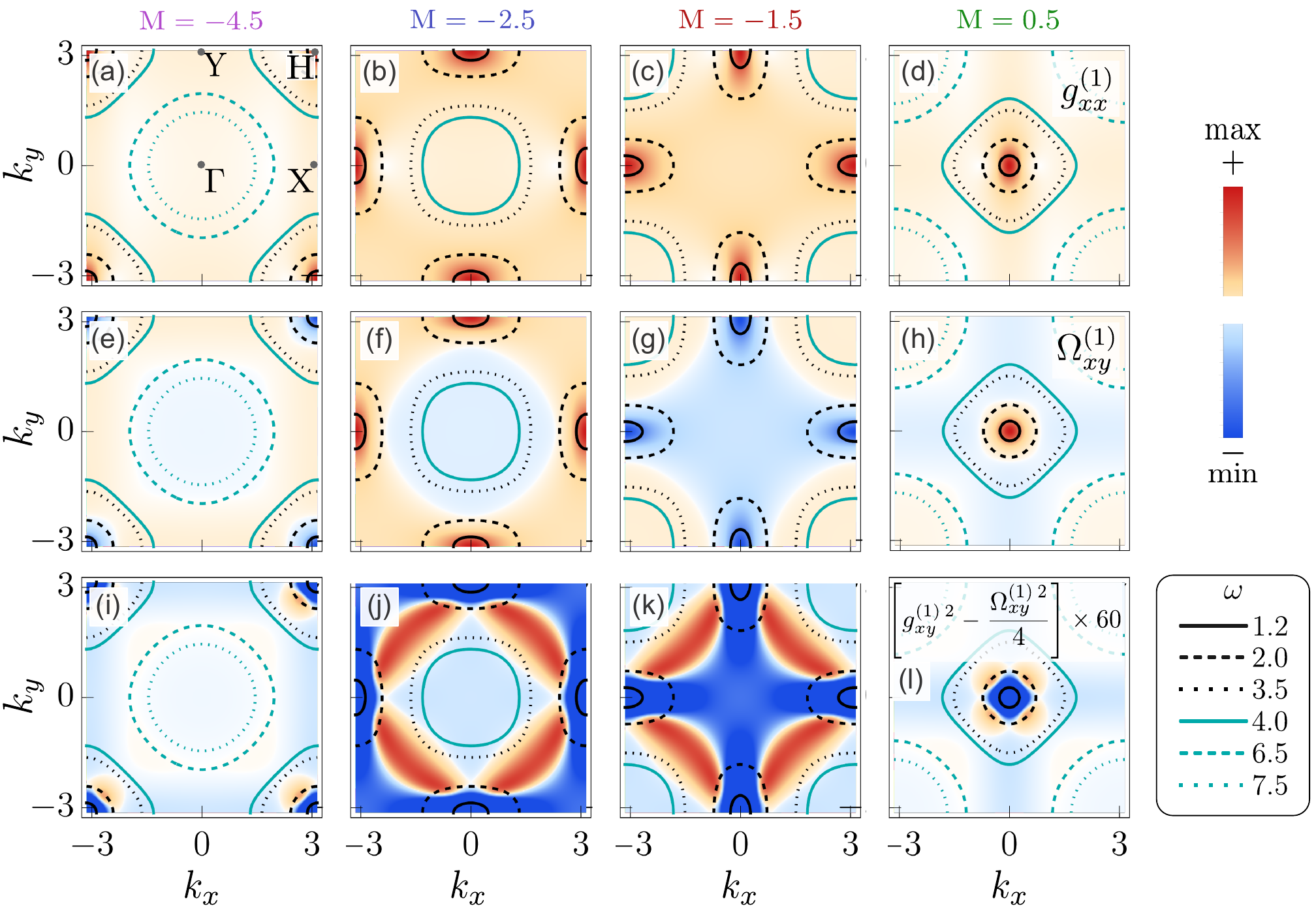}
    \caption{Quantum geometry of a Chern insulator on a square lattice. The panels show the $g_{xx}^{(1)}(\mathbf{k})$ component of the quantum metric (first row), the Berry curvature $\Omega_{xy}^{(1)}(\mathbf{k})$ (second row), and the combination $\left[g_{xy}^{(1)}(\mathbf{k})\right]^2-\left[\frac{\Omega_{xy}^{(1)}(\mathbf{k})}{2}\right]^2$ (third row) for the valence band and for the same values of $M$ as in Fig.\ref{fig:linear_square}. Each quantity is normalized by its maximum absolute value in momentum space, except for $\left[g_{xy}^{(1)}(\mathbf{k})\right]^2-\left[\Omega_{xy}^{(1)}(\mathbf{k})/2\right]^2$, which is multiplied by $60$ after normalization to improve the visibility of its momentum-space sign structure. The high-symmetry points in the Brillouin zone $\Gamma$ ($\mathbf{k=0}$), $X$ ($\mathbf{k}=\frac{\mathbf{b}_1}{2}$), $Y$ ($\mathbf{k}=\frac{\mathbf{b}_2}{2}$), and $H$ ($\mathbf{k}=\frac{\mathbf{b}_1+\mathbf{b}_2}{2}$), where $\mathbf{b}_j$ denote the primitive vectors of the reciprocal lattice, are indicated in panel (a).}  
\label{fig:quantum_geometry_square_part1}
\end{figure*}

A useful symmetry emerges about $M_{c,2}=-2$. The band dispersions for $M=M_{c,2}+\delta$ and $M=M_{c,2}-\delta$, with $\delta>0$ are identical, so that the corresponding spectral features occur at the same frequencies. Their geometric weights, however, transform differently: the OAR-integrated quantum metric is unchanged, whereas the Berry curvature changes its overall sign. Consequently, $\chi_{xx}^{(1)}(\omega)$ is identical for masses equidistant from $M_{c,2}$, while $\chi_{xy}^{(1)}(\omega)$ changes sign, as shown in Fig.\ref{fig:linear_square}. The momentum-space profiles of the Berry curvature and quantum metric for different values of $M$ are shown in Fig.\ref{fig:quantum_geometry_square_part1}.

Next, we turn to the 2DCS of this system. As in the continuum Dirac model, spatial inversion enforces $\chi^{(2)}(\omega_1,\omega_2)=0$, and and we focus on third-order ABB response. We discuss in the following the paramagnetic contribution to the ABB response, while the corresponding contact-term contribution is shown in the Appendix \ref{app:JDOS_square}. Figure \ref{fig:2DCS_square} shows the $xxyx$ and $xyyx$ components for representative values of $M$. The symmetry about $M_{c,2}=-2$ is again apparent. The $xyyx$ spectra are identical for masses equidistant from $M_{c,2}$, while the $xxyx$ spectra show an overall sign change. 

At frequencies close to the minimum interband gap, the lattice spectra resemble those of the continuum Dirac model because the dominant transitions involve states near the gap minima. Lattice-specific features emerge at higher frequencies, as the resonant contours explore larger portions of the Brillouin zone. In particular, additional peaks appear along the main diagonal at frequencies where the JDOS is singular, accompanied by sharp features along the horizontal axis and the anti-diagonal.

As established in the continuum model, the echo signal probes the OAR-integrated $g_{xx}^{(1)}(\mathbf{k})\Omega_{xy}^{(1)}(\mathbf{k})$ in the $xxyx$ component and $\left[g_{xy}^{(1)}(\mathbf{k})\right]^2-\left[\frac{\Omega_{xy}^{(1)}(\mathbf{k})}{2}\right]^2$ in the $xyyx$ component. In the lattice model, the corresponding OAR-integrated quantities may change sign multiple times as the resonant contours of the OAR evolve across the Brillouin zone, resulting in nodes in the echo signal along $\omega_2=-\omega_1$.

For the $xxyx$ echo, we find a distinct parity of the node count in the trivial and topological regimes, as shown in Fig.\ref{fig:2DCS_square}(a)-(d). Consider first $M$ near $M_{c,1}=0$. In the trivial regime, with $M>0$, the low-frequency response is dominated by states around $\Gamma$ ($\mathbf{k}=0$), where $g_{xx}^{(1)}(\mathbf{k})\Omega_{xy}^{(1)}(\mathbf{k})$ is positive. As a result, the corresponding OAR-integrated geometric weight is positive. As $\omega$ increases, the OAR moves in the Brillouin zone and, for high frequencies, the resonant contour is centered around $H$ ($\mathbf{k}=\frac{\mathbf{b}_1+\mathbf{b}_2}{2}$), where the same geometric quantity is also positive. Assuming a continuous evolution of the echo amplitude with frequency, equal signs at the two spectral edges constrain the number of sign-changing nodes to be even. For $M=0.5$, we indeed find two nodes.

In the topologically non-trivial regime near $M_{c,2}=-2$, the low-frequency contribution originates from the valleys $X$ ($\mathbf{k}=\mathbf{b}_1/2$) and $Y$ ($\mathbf{k}=\mathbf{b}_2/2$) and is negative, while the high-frequency contribution around $H$ is positive [see Fig.\ref{fig:quantum_geometry_square_part1}(g)]. Therefore, the echo must change sign an odd number of times. In fact, for $M=-1.5$, we find one node in the echo signal. Given the symmetry around $M_{c,2}=-2$, similar arguments apply for $-4<M<-2$, with the roles of $\Gamma$ and $H$ interchanged. More generally, within this model the signs of the geometric weight at the lower and upper spectral edges constrain the parity of the $xxyx$ echo-node count to be even for $C=0$ and odd for $|C|=1$. This distinction does not resolve the sign of the Chern number.

The $xyyx$ echo is controlled by the OAR integral of $\left[g_{xy}^{(1)}(\mathbf{k})\right]^2-\left[\frac{\Omega_{xy}^{(1)}(\mathbf{k})}{2}\right]^2$. Its nodes likewise occur when this integral changes sign. In this case, however, the node count is not tied in any simple way to the Chern number. At the high-symmetry points $\Gamma$, $X$, $Y$, and $H$, and away from the critical masses, the corresponding geometric quantity is negative at both spectral edges, constraining sign-changing nodes to appear or disappear in pairs. Consequently, the number of $xyyx$ echo nodes is even but can vary with $M$ within a single Chern phase(see Appendix \ref{app:JDOS_square}).

\begin{figure*}[t!]
    \centering
    \includegraphics[width=0.99\linewidth]{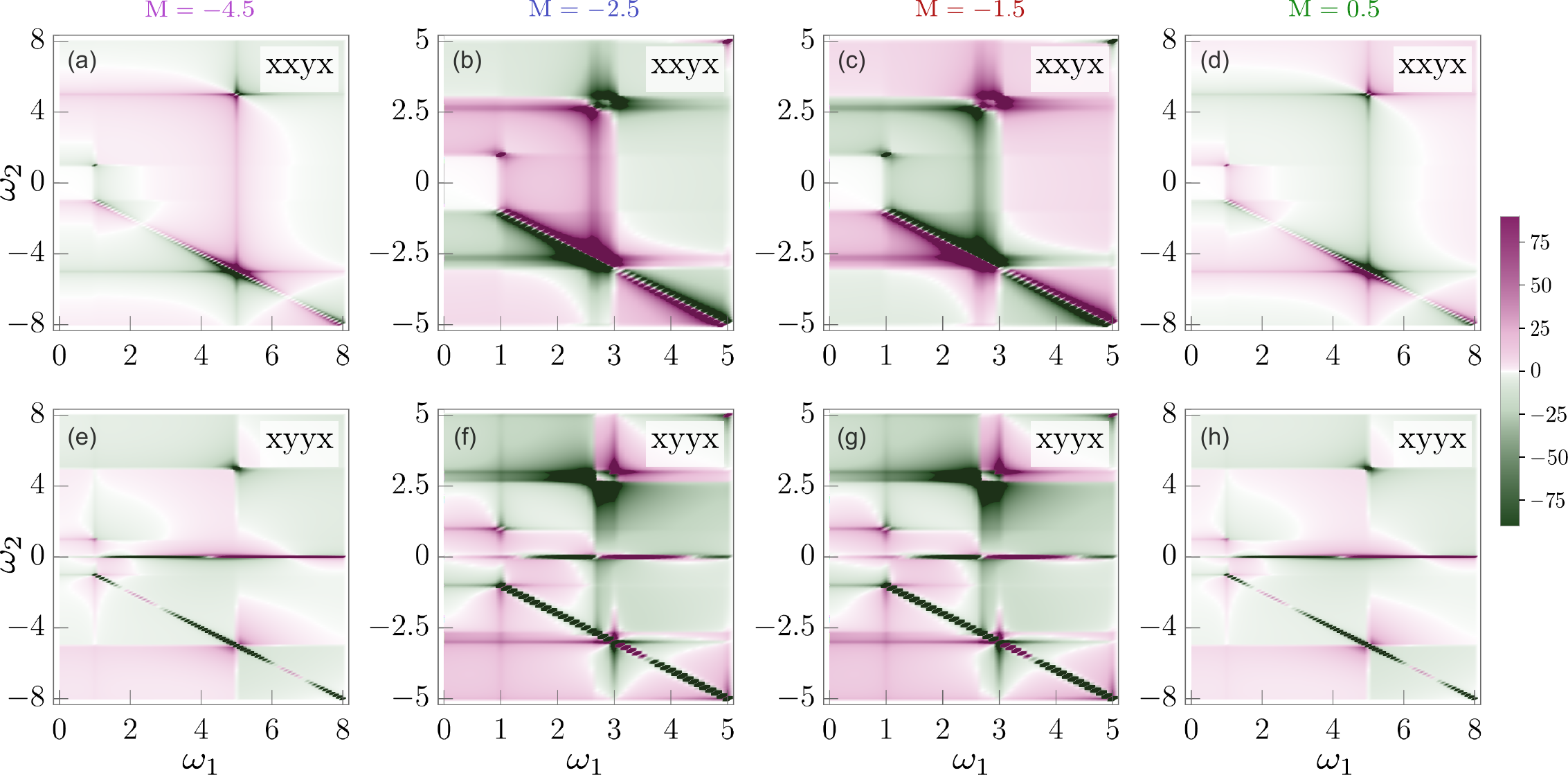}
    \caption{Two-dimensional ABB spectra of a square-lattice Chern insulator. The first and second rows show the imaginary part of $4\pi^2\chi_{xxyx}^{\text{(ABB)}}(\omega_1,\omega_2)$ and $4\pi^2\chi_{xyyx}^{\text{(ABB)}}(\omega_1,\omega_2)$, respectively, for the same values of $M$ as in Fig.\ref{fig:linear_square}.}
\label{fig:2DCS_square}
\end{figure*}

\section{Conclusions}

In this work, we developed a quantum geometric formulation of multidimensional optical susceptibilities in insulating systems. Working in the velocity gauge, we derived analytical expressions for the second-order and third-order 2DCS. Although we focused on the lowest nonlinear orders, the structure of the derivation provides a systematic route to higher-order multidimensional responses.

Our results show that different regions in the frequency plane of the multidimensional optical spectra probe distinct quantum-geometric quantities. The second-order 2DCS, $\chi^{(2)}(\omega_1,\omega_2)$, has stronger spectroscopic features on the diagonal and horizontal axis of the frequency plane. These features probe different Cartesian components of the relative-velocity-weighted pair-resolved quantum geometric tensor. In the two-band limit, the pair-resolved tensor reduces to the quantum geometric tensor of the occupied band, and the corresponding frequency-resolved quantities are related to dipoles of the quantum metric and Berry curvature. The second-order 2DCS also supports additional multiband off-diagonal features, which probe geometric quantities constructed from trilinear products of non-Abelian Berry connection matrix elements.

One of the central result of this work is the quantum-geometric content of the third-order ABB 2DCS. Unlike the second-order susceptibility, the ABB response remains finite in centrosymmetric systems, for which $\chi^{(2)}(\omega_1,\omega_2)$ vanishes identically because of spatial inversion symmetry. As in the second-order response, distinct regions of the ABB spectrum probe different geometric structures beyond the conventional quantum metric and Berry curvature. We focused primarily on the paramagnetic contribution to the ABB response, which supports an echo signal along the anti-diagonal, $\omega_2=-\omega_1$. This echo probes gauge-invariant products of four interband Berry-connection matrix elements. In the two-band limit, these products reduce to bilinear combinations of components of the quantum geometric tensor of the occupied band. In multiband systems, additional spectroscopic features arise from geometric structures absent in the two-band limit, including relative-velocity-weighted trilinear products of interband Berry connections. We also analyzed the contact contribution separately and showed that it produces additional resonant structures along the main diagonal and the horizontal axis of the frequency plane with a distinct geometric content. 

We illustrated these results using minimal models for two-dimensional Chern insulators. In both the continuum Dirac model and its square-lattice regularization, the ABB echo reflects the momentum space sign structure of the geometric quantity selected by the tensor component as the resonant contour evolves through momentum space. Interestingly, for the $xxyx$ response of the models considered here, we found a simple relation between the echo-node count and topology. In the continuum model, the echo has one node in the topologically trivial regime, reflecting the sign change of the Berry curvature in momentum space. For the lattice model, it is the parity of the echo-node count that changes between the topologically trivial and non-trivial regime. We found the number of sign-changing nodes is even in the topologically trivial regime ($C=0$), and odd in the non-trivial regime ($\left|C\right|=1$). By contrast, the nodal structure of the $xyyx$ echo is not as directly related to topology. Instead, its nodes reflect cancellations between momentum-space regions where $\left[g_{xy}^{(1)}\right]^2-\left[\frac{\Omega_{xy}^{(1)}}{2}\right]^2$ has opposite signs, and their number can vary within the same Chern phase.

Multidimensional spectroscopy is a rapidly developing field in condensed matter physics. We propose multidimensional coherent spectroscopy as a versatile probe of quantum geometry, capable not only of detecting geometric effects but also of separating distinct geometric structures in frequency space. Our work demonstrates that multidimensional spectroscopy can provide information about the quantum geometry of the electrons beyond that available from linear responses. We focused on the third-order 2D responses of clean insulators described within the tight-binding description. Future research directions include the role of intraband processes in higher-order responses of metals and semimetals, the effect of disorder, and the signatures of electronic interactions in the 2DCS. We hope that these results will motivate further theoretical and experimental investigations of multidimensional spectroscopic probes of quantum geometry.

\section{Acknowledgments}
T.V.T. acknowledges financial support from the CNPq grant 312329/2023-7 and the University of S\~{a}o Paulo startup grant number 22.1.09345.01.2. M.P. was funded in part by The Robert A. Welch Foundation, Grant \#L-E-0001-19921203.

\bibliographystyle{apsrev4-2}
\bibliography{References}
\appendix
%--------------------------

\section{Optical susceptibility tensors in terms of current-operator matrix elements}\label{app:chi_J}

As mentioned in the main text, the optical susceptibility tensors expressed in the velocity gauge have a hierarchy of terms involving linear and higher-order derivatives of the Bloch Hamiltonian, which are generically called multi-photon vertices in the literature~\cite{parker2019diagrammatic,watanabe2020general}. In this section, we write the full expressions for the linear ($n=1$), second-order ($n=2$), and third-order ($n=3$) optical response functions and explain which multi-photon vertices contribute to the multidimensional spectra when $n\geq 2$. The analysis presented here can be readily extended to higher-order response functions and higher-order multidimensional spectra.

We start from the light-induced current density, 
\begin{equation}
    J_{\mu}(t)=\frac{1}{V}\text{tr}\left(\hat{\rho}_{I}(t)\hat{J}_{\mu}\right) \text{ ,}
    \label{eq:J_average}
\end{equation}

\noindent where $V$ is the system volume (or area, in the case of a 2D model), $\hat{J}_{\mu}$ is the current operator and $\hat{\rho}_{I}(t)$ is the system density operator in the interaction picture. Its equation of motion is 
\begin{equation}
    \frac{d\hat{\rho}_{I}(t)}{dt}=-\frac{i}{\hbar}\left[\hat{H}_{\text{1,I}}(t),\hat{\rho}_I(t)\right] \text{ .}
    \label{eq:EOM_rho}
\end{equation}

\noindent Here, $\hat{H}_{\text{1,I}}(t)=e^{i\hat{H}_0t/\hbar}\hat{H}_1e^{-i\hat{H}_0t/\hbar}$ is the Hamiltonian describing the light-matter interaction, also in the interaction picture. Besides, as in the main text,  $\hat{H}_0=\sum\limits_{\mathbf{k}}\hat{\psi}_{\mathbf{k}}^{\dag}\hat{h}(\mathbf{k})\hat{\psi}^{\null}_{\mathbf{k}}$ denotes the Hamiltonian of the system in the absence of the light. The first-quantized Bloch Hamiltonian $\hat{h}(\mathbf{k})$ is kept generic here. 

In the velocity gauge, the coupling between system and light is given by the Peierls substitution~\cite{peierls1933theorie,graf1995electromagnetic},
\begin{equation}
    \hat{H}[\mathbf{A}(t)]=\sum\limits_{\mathbf{k}}\hat{\psi}_{\mathbf{k}}^{\dag}\hat{h}\left(\mathbf{k}+\frac{e}{c\hbar}\mathbf{A}(t)\right)\hat{\psi}^{\null}_{\mathbf{k}} \text{ ,}
\end{equation}

\noindent where, $e>0$ is the fundamental charge, $c$ is the speed of light, and $\mathbf{A}(t)$ is the vector potential of the light-field in the dipole approximation. For the pulse sequence we consider in this work, $\mathbf{A}(t)=\mathbf{A}_1\delta(t)+\mathbf{A}_2\delta(t-\tau)$, but in this section we keep $\mathbf{A}(t)$ generic. Expanding $\hat{H}[\mathbf{A}(t)]$ in powers of the vector potential yields $\hat{H}[\mathbf{A}(t)]=\hat{H}_0+\hat{H}_{1}(t)$, with 
\begin{align}
    \hat{H}_1(t)&=-\frac{1}{c}\left[\hat{J}_{0,\mu}A_{\mu}(t)+\frac{1}{2}\hat{J}_{1,\mu\nu}A_{\mu}(t)A_{\nu}(t)\right.\nonumber\\
    &\left.+\frac{1}{3!}\hat{J}_{2,\mu\nu\gamma}A_{\mu}(t)A_{\nu}(t)A_{\gamma}(t)+\mathcal{O}(A^4)\right] \text{ .}
    \label{eq:H1}
\end{align}

\noindent Here Einstein notation is implicit, and $\hat{J}_{n}$ (with $n=0,1,2,\cdots$) are the components entering the expansion of the full current operator in powers of the vector potential, 
\begin{equation}
    \hat{J}_{\mu}=-c\frac{\partial H[\mathbf{A}(t)]}{\partial A_{\mu}}=\hat{J}_{0,\mu}+\hat{J}_{1,\mu\nu}A_{\nu}(t)+\mathcal{O}(A^2) \text{ .}
    \label{eq:Jop_full}
\end{equation}

\noindent Following the literature, we call $\hat{J}_n$ the $(n+1)$-photon vertex~\cite{parker2019diagrammatic}.

The operator $\hat{J}_{0,\mu}=(-c)\left.\frac{\partial H[\mathbf{A}]}{\partial A_{\mu}}\right|_{\mathbf{A}=0}$ in the first term on the right-hand side of Eq.(\ref{eq:H1}) is the paramagnetic part of the current operator or, equivalently, single-photon current vertex. Its expression is given in Eq.(\ref{eq:J_orbital}) in the main text. Moreover, in the band basis, its matrix elements $j_{nm}^{(\mu)}(\mathbf{k})$ correspond to the covariant derivative of the Bloch Hamiltonian [see Eq.(\ref{eq:j_Bloch})]. The next term on the right-hand side of Eq.(\ref{eq:H1}) is quadratic in $\mathbf{A}(t)$. Therefore, $\hat{J}_{1,\mu\nu}=(-c)\left.\frac{\partial^2 H[\mathbf{A}]}{\partial A_{\mu}\partial A_\nu}\right|_{\mathbf{A}=0}$ is also called the diamagnetic current vertex. Its expression is also given in the main text [see Eq.(\ref{eq:J1_orbital})] and its matrix elements $s_{nm}^{(\mu\nu)}(\mathbf{k})$ in the band basis are the covariant derivative of $j_{nm}^{(\mu)}(\mathbf{k})$. Using Eq.(\ref{eq:j_Bloch}), we can express $s_{nm}^{(\mu\nu)}(\mathbf{k})$ in terms of the non-Abelian Berry connection, 
\begin{align}
    &s_{nm}^{(\mu\nu)}(\mathbf{k})=
(\partial_\mu \partial_\nu \varepsilon_n(\mathbf{k}))\delta_{nm}
+i(\partial_\mu \Delta_{nm})\mathcal{A}_{nm}^{(\nu)}(\mathbf{k})
\nonumber\\[0.2cm]
&+i(\partial_\nu \Delta_{nm})\mathcal{A}_{nm}^{(\mu)}(\mathbf{k})+i\frac{\Delta_{nm}}{2}\left(
\partial_\mu \mathcal{A}_{nm}^{(\nu)}(\mathbf{k})+
\partial_\nu \mathcal{A}_{nm}^{(\mu)}(\mathbf{k})
\right)\nonumber\\[0.2cm]
&+\frac{1}{2}\sum_p
\left(\Delta_{pm}-\Delta_{np}\right)\left(
\mathcal{A}_{np}^{(\mu)}\mathcal{A}_{pm}^{(\nu)}+
\mathcal{A}_{np}^{(\nu)}\mathcal{A}_{pm}^{(\mu)}
\right).
\label{eq:snm_A}
\end{align}

\noindent This pattern generalizes for higher-order terms: the matrix elements of $\hat{J}_n$ in band basis is the covariant derivative of the matrix elements of $\hat{J}_{n-1}$. We recall that $\Delta_{nm}(\mathbf{k})=\varepsilon_{n}(\mathbf{k})-\varepsilon_{m}(\mathbf{k})$ and $\mathcal{A}_{nm}^{(\mu)}(\mathbf{k})=i\left\langle \left.n(\mathbf{k})\right|\partial_{\mu}m(\mathbf{k})\right\rangle$.

We now have all ingredients to obtain the optical responses at any order in powers of $\mathbf{A}(t)$. Substituting Eq.(\ref{eq:H1}) in Eq.(\ref{eq:EOM_rho}), solving iteratively for $\rho_{I}(t)$ and substituting the result, together with Eq.(\ref{eq:Jop_full}) in Eq.(\ref{eq:J_average}) yields, in linear order in $\mathbf{A}$ the following: 
\begin{equation}
    J_{\mu}^{(1)}(t)=\int\limits_{-\infty}^{\infty}ds\,K_{\mu\nu}^{(1)}(s)A_{\nu}(t-s)\text{ ,}
\end{equation}

\noindent where
\begin{equation}
    VK_{\mu\nu}^{(1)}(s)=\delta(s)\left\langle\hat{J}_{1,\mu\nu}\right\rangle+\frac{i}{\hbar c}\theta(s)\left\langle\left[\hat{J}_{0,\mu}(s),\hat{J}_{0,\nu}(0)\right]\right\rangle\text{ .}
    \label{eq:K1}
\end{equation}

\noindent Here, $\left\langle\cdots\right\rangle=\text{tr}\left(\cdots\hat{\rho_0}\right)$, and $\hat{\rho}_0$ is the ground state equilibrium density matrix. Besides. For a detailed step-by-step derivation of these expressions using perturbation theory, we refer the reader to Refs.~\cite{parker2019diagrammatic,watanabe2020general}. For compactness, we present the response in the form $V K^{(1)}$, where $V$ is the system volume. We will keep this notation for the higher-order responses.

The second term in the right-hand side of Eq.(\ref{eq:K1}), which involves only $\hat{J}_0$, is the paramagnetic part of the linear response. As shown in the main text, it carries information about the quantum geometric tensor and in this work it was calculated for topological insulators [see Fig.\ref{fig:Linear_untilted} and Fig.\ref{fig:linear_square}]. Furthermore, the first term in the right-hand side of Eq.(\ref{eq:K1}) is a constant in the frequency space, 
\begin{equation}
   \left\langle\hat{J}_{1,\mu\nu}\right\rangle=-\frac{e^2}{\hbar^2c}\sum\limits_{n,\mathbf{k}} f_{n}(\mathbf{k})s_{nn}^{(\mu\nu)}(\mathbf{k}) \text{ .}
   \label{eq:contact_1}
\end{equation}

\begin{figure}[t!]
    \centering
    \includegraphics[width=0.99\linewidth]{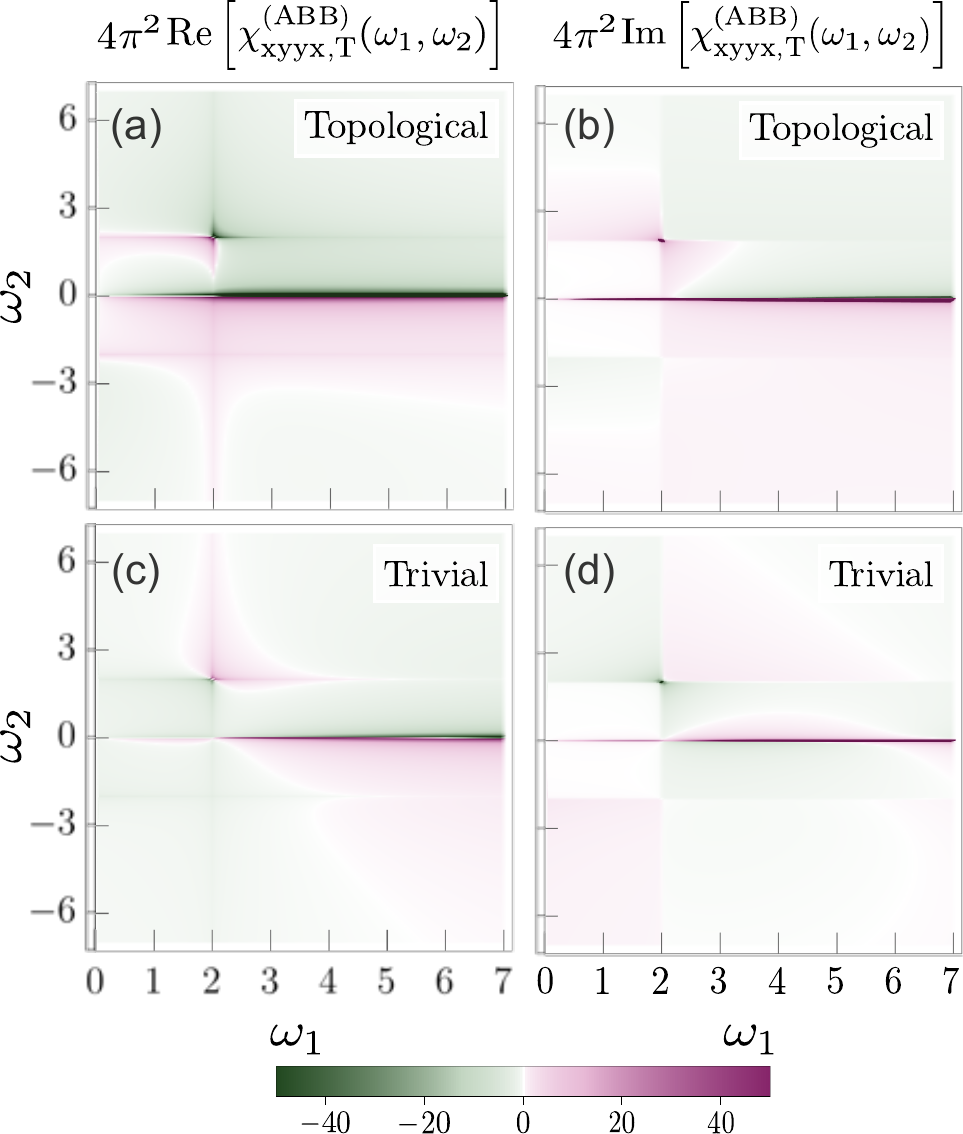}
    \caption{Contact term component of the ABB $xyyx$ response of the continuum Dirac model considered in Sec.\ref{sec:Dirac}. We set $B/M=0.2$ in the upper row (topological regime) and $B/M=-0.2$ in the lower row (trivial regime).}
    \label{fig:contact_continnum}
\end{figure}

\noindent Note that the diagonal element
\begin{equation}
    s_{nn}^{(\mu\nu)}(\mathbf{k})=
\partial_\mu\partial_\nu\varepsilon_n(\mathbf{k})-
2\sum_{p\neq n}
\Delta_{np}(\mathbf{k})
\operatorname{Re}\left[
\mathcal A_{np}^{(\mu)}(\mathbf{k})
\mathcal A_{pn}^{(\nu)}(\mathbf{k})
\right].
\label{eq:snn}
\end{equation}
\noindent contains a band-curvature term (first term in the right-hand side of Eq.(\ref{eq:snn})) and a geometric interband term. The band-curvature part of Eq.(\ref{eq:contact_1}) can be rewritten, via integration by parts, as an intraband Fermi surface contribution, which vanishes in the zero-temperature response of insulators we consider in this work. The remaining interband contribution is generically finite in insulators. Since the $\hat{J}_1$ term enters the time-domain response through a contact term proportional to $\delta(s)$, its contribution to the response function is frequency independent and therefore appears as a constant contribution alongside the frequency-dependent paramagnetic response. To emphasize the frequency-dependent resonant features in the linear responses calculated in this work, we subtract this constant contact contribution from the displayed linear-response spectra.

In second order in the vector potential, we obtain  
\begin{equation}
    J_{\mu}^{(2)}(t)=\int\limits_{-\infty}^{\infty}ds_1\int\limits_{-\infty}^{\infty}ds_2\,K_{\mu\nu\gamma}^{(2)}(s_1,s_2)A_{\nu}(t-s_1)A_{\gamma}(t-s_2)\text{ ,}
\end{equation}

\noindent with
\begin{widetext}
\begin{align}
VK_{\mu\nu\gamma}^{(2)}(s_1,s_2)=& \frac{1}{2}\delta(s_2)\delta(s_1)\left\langle
\hat{J}_{2,\mu\nu\gamma}\right\rangle
+\left(\frac{i}{\hbar c}\right)^2
\theta(s_1)\theta(s_2-s_1)
\left\langle\left[\left[
\hat{J}_{0,\mu}(s_2),\hat{J}_{0,\nu}(s_2-s_1)\right],\hat{J}_{0,\gamma}(0)\right]\right\rangle\nonumber\\[0.2cm]
&+
\frac{i}{2\hbar c}
\delta(s_2-s_1)\theta(s_1)
\left\langle
\left[
\hat{J}_{0,\mu}(s_1),
\hat{J}_{1,\nu\gamma}(0)\right]\right\rangle
+
\frac{i}{\hbar c}\delta(s_1)\theta(s_2)
\left\langle\left[
\hat{J}_{1,\mu\nu}(s_2),\hat{J}_{0,\gamma}(0)\right]\right\rangle
\text{ .}
\label{eq:K2}
\end{align}
\end{widetext}

\noindent Here, $s_1$ and $s_2$ are connected with the time intervals $\tau_1$ and $\tau_2$ defined in the main text [see Fig.\ref{fig:setup_2p}] via $\tau_1=s_2-s_1$ and $\tau_2=s_1$. The first term in the right-hand side of Eq.(\ref{eq:K2}) is a three-photon vertex term. Similarly to the term proportional to $\left\langle\hat{J}_{1,\mu\nu}\right\rangle$ in $\mathbf{K}_{\mu\nu}^{(1)}$, it is a constant term in the frequency space. However, it does not contribute to the second-order 2DCS. That is because the three-photon vertex requires the two pulses in the pulse sequence to overlap, which does not happen since we keep $\tau_1>0$ and $\tau_2>0$. Indeed, the three-photon vertex in $K^{(2)}$ is eliminated by the subtraction $\mathbf{J}_{NL}(t)=\mathbf{J}_{12}(t)-\mathbf{J}_{1}(t)-\mathbf{J}_{2}(t)$ that yield the purely non-linear light-induced current density that is relevant for the 2DCS. Likewise, the third and fourth terms in the right-hand side of $K^{(2)}$ are also absent in the 2DCS due to the finite time interval between the pulses and between pulses and measurement.

Therefore, only the paramagnetic term (second term in the right-hand side of $K^{(2)}$) contributes to the 2DCS response, which reads, in the frequency space,  
\begin{widetext}
\begin{equation}
 \chi_{\mu\nu\gamma}^{(2)}(\omega_1,\omega_2)=-\frac{1}{V}\sum\limits_{\mathbf{k},n, m, l}f_{nl}(\mathbf{k})\left[\frac{j_{nm}^{(\mu)}(\mathbf{k})j_{ml}^{(\nu)}(\mathbf{k})j_{ln}^{(\gamma)}(\mathbf{k})}{\left(\omega_1-\Delta_{ln}(\mathbf{k})+i\Gamma\right)\left(\omega_2-\Delta_{mn}(\mathbf{k})+i\Gamma\right)}+\frac{j_{mn}^{(\mu)}(\mathbf{k})j_{lm}^{(\nu)}(\mathbf{k})j_{nl}^{(\gamma)}(\mathbf{k})}{\left(\omega_1+\Delta_{ln}(\mathbf{k})+i\Gamma\right)\left(\omega_2+\Delta_{mn}(\mathbf{k})+i\Gamma\right)}\right] \text{ .}
 \label{eq:chi_2}
\end{equation}
\end{widetext}

\noindent We emphasize that $\omega_1$ and $\omega_2$ are the frequencies conjugates to $\tau_1$ and $\tau_2$, respectively. Splitting the sum over the bands into the $\varepsilon_l(\mathbf{k})>\varepsilon_{n}(\mathbf{k})$ and  $\varepsilon_l(\mathbf{k})<\varepsilon_{n}(\mathbf{k})$ contributions and making a change of variables $(n,l)\rightarrow(l,n)$ in the latter, we obtain
\begin{align}
    \chi_{\mu\nu\gamma}^{(2)}(\omega_1,\omega_2)=-\sum\limits_{\substack{\mathbf{k},m,n,l \\ \varepsilon_l(\mathbf{k})>\varepsilon_n(\mathbf{k})}}\frac{f_{nl}(\mathbf{k})}{V}&\left[L_{\mu\nu\gamma}^{(2;a)}(\omega_1,\omega_2)\right.\nonumber\\
&+\left.L_{\mu\nu\gamma}^{(2;b)}(\omega_1,\omega_2)\right]\text{ ,}
\end{align}

\noindent where
\begin{align}
    L_{\mu\nu\gamma}^{(2;a)}(\omega_1,\omega_2)&=\frac{j_{nm}^{(\mu)}(\mathbf{k})j_{ml}^{(\nu)}(\mathbf{k})j_{ln}^{(\gamma)}(\mathbf{k})}{\left(\omega_1-\Delta_{ln}(\mathbf{k})+i\Gamma\right)\left(\omega_2-\Delta_{mn}(\mathbf{k})+i\Gamma\right)}\nonumber\\[0.2cm]
    &-\frac{j_{ml}^{(\mu)}(\mathbf{k})j_{nm}^{(\nu)}(\mathbf{k})j_{ln}^{(\gamma)}(\mathbf{k})}{\left(\omega_1-\Delta_{ln}(\mathbf{k})+i\Gamma\right)\left(\omega_2+\Delta_{ml}(\mathbf{k})+i\Gamma\right)}\text{ ,}
    \label{eq:L2a}
\end{align}

\noindent and $L_{\mu\nu\gamma}^{(2;b)}(\omega_1,\omega_2)=\left[L_{\mu\nu\gamma}^{(2;a)}(-\omega_1,-\omega_2)\right]^*$. Note that $L_{\mu\nu\gamma}^{(2;a)}$ generates resonant  spectral features only on the half-frequency plane $\omega_1>0$. Substituting Eq.(\ref{eq:j_Bloch}) into Eq.(\ref{eq:L2a}), and reorganizing the terms into $m=n$, $m=l$ and $m\neq n,l$ parts, we obtain $\chi_{\mu\nu\gamma,+}^{(2)}(\omega_1,\omega_2)$ defined in Eq.(\ref{eq:chi_2main}) in the main text.  

We now focus in the third-order response, 
\begin{equation}
    J_{\mu}^{(3)}(t)=\int\limits_{-\infty}^{\infty}\prod\limits_{i=1}^{3}ds_i\,K_{\mu\nu\gamma\delta}^{(3)}A_{\nu}(t-s_1)A_{\gamma}(t-s_2)A_{\delta}(t-s_3)\text{ .}
\end{equation}

\noindent The time-dependence of the response function $K_{\mu\nu\gamma\delta}^{(3)}(s_1,s_2,s_3)$ was omitted to shorten the notation. It has eight distinct terms that account for all possible combinations of single-photon, two-photon, and three-photon vertices:
\begin{widetext}
\begin{align}
&VK_{\mu\nu\gamma\delta}^{(3)}(s_1,s_2,s_3)= \frac{1}{3!}\delta(s_1)\delta(s_2)\delta(s_3)\left\langle
\hat{J}_{3,\mu\nu\gamma\delta}\right\rangle+\frac{i}{2\hbar c}\delta(s_1)\delta(s_2)\theta(s_3)\left\langle\left[\hat{J}_{2,\mu\nu\gamma}(s_3),\hat{J}_{0,\delta}(0)\right]\right\rangle
\nonumber\\[0.1cm]
&+\frac{i}{2\hbar c}\delta(s_3-s_2)\delta(s_1)\theta(s_2)\left\langle\left[\hat{J}_{1,\mu\nu}(s_3),\hat{J}_{1,\gamma\delta}(0)\right]\right\rangle+\frac{i}{3!\hbar c}\delta(s_3-s_2)\delta(s_3-s_1)\theta(s_3)\left\langle\left[\hat{J}_{0,\mu}(s_3),\hat{J}_{2,\nu\gamma\delta}(0)\right]\right\rangle\nonumber\\[0.1cm]
&+
\frac{1}{2}\left(\frac{i}{\hbar c}\right)^2
\delta(s_3-s_2)\theta(s_3-s_1)\theta(s_1)
\left\langle
\left[\left[
\hat{J}_{0,\mu}(s_3),
\hat{J}_{0,\nu}(s_3-s_1)\right],\hat{J}_{1,\gamma\delta}(0)\right]\right\rangle\nonumber\\[0.1cm]
&+\left(\frac{i}{\hbar c}\right)^2
\delta(s_1)\theta(s_3-s_2)\theta(s_2)
\left\langle\left[\left[
\hat{J}_{1,\mu\nu}(s_3),\hat{J}_{0,\gamma}(s_3-s_2)\right],\hat{J}_{0,\delta}(0)\right]\right\rangle\nonumber\\[0.1cm]
&+\frac{1}{2}\left(\frac{i}{\hbar c}\right)^2
\delta(s_2-s_1)\theta(s_3-s_2)\theta(s_2)
\left\langle\left[\left[
\hat{J}_{0,\mu}(s_3),\hat{J}_{1,\nu\gamma}(s_3-s_2)\right],\hat{J}_{0,\delta}(0)\right]\right\rangle\nonumber\\[0.1cm]
&+\left(\frac{i}{\hbar c}\right)^3\theta(s_3-s_2)\theta(s_2-s_1)\theta(s_1)\left\langle\left[\left[\left[
\hat{J}_{0,\mu}(s_3),\hat{J}_{0,\nu}(s_3-s_1)\right],\hat{J}_{0,\gamma}(s_3-s_2)\right],\hat{J}_{\delta}(0)\right]\right\rangle
\text{ .}
\label{eq:K3}
\end{align}
\end{widetext}

\noindent Here, $s_1$, $s_2$ and $s_3$ are related to the time intervals defined in the main text [see Fig.\ref{fig:pulse_setup}(a)] via $t_1=s_3-s_2$, $t_2=s_2-s_1$ and $t_3=s_1$. Similarly to the case of $K^{(2)}$, not all terms of the previous equation contributes to the multidimensional spectra. For instance, only the last term in the right-hand side of $K^{(3)}$ contributes for the 3DCS, since the remaining terms require overlap of pulses in the sequence and are either eliminated in the subtraction $\mathbf{J}_{NL}(t)=\mathbf{J}_{12}(t)-\mathbf{J}_{1}(t)-\mathbf{J}_{2}(t)$ or are inconsistent with the time order of the pulses and subsequent measurement. Therefore, the 3DCS has purely paramagnetic contribution, which in frequency space, reads 
\begin{widetext}
\begin{align}
    \chi_{\mu\nu\gamma\delta}^{(3)}(\omega_1,\omega_2,\omega_3)=-\frac{1}{V}\sum\limits_{\mathbf{k}}\sum\limits_{m,n,p,s}f_{ns}&(\mathbf{k})\left[\frac{j_{nm}^{(\mu)}(\mathbf{k})j_{mp}^{(\nu)}(\mathbf{k})j_{ps}^{(\gamma)}(\mathbf{k})j_{sn}^{(\delta)}(\mathbf{k})}{\left(\omega_1-\Delta_{sn}(\mathbf{k})+i\Gamma\right)\left(\omega_2-\Delta_{pn}(\mathbf{k})+i\Gamma\right)\left(\omega_3-\Delta_{mn}(\mathbf{k})+i\Gamma\right)}\right.\nonumber\\[0.2cm]
    &-\frac{j_{mp}^{(\mu)}(\mathbf{k})j_{nm}^{(\nu)}(\mathbf{k})j_{ps}^{(\gamma)}(\mathbf{k})j_{sn}^{(\delta)}(\mathbf{k})}{\left(\omega_1-\Delta_{sn}(\mathbf{k})+i\Gamma\right)\left(\omega_2-\Delta_{pn}(\mathbf{k})+i\Gamma\right)\left(\omega_3-\Delta_{pm}(\mathbf{k})+i\Gamma\right)}\nonumber\\[0.2cm]
    &\left.-\frac{j_{mn}^{(\mu)}(\mathbf{k})j_{pm}^{(\nu)}(\mathbf{k})j_{sp}^{(\gamma)}(\mathbf{k})j_{ns}^{(\delta)}(\mathbf{k})}{\left(\omega_1-\Delta_{ns}(\mathbf{k})+i\Gamma\right)\left(\omega_2-\Delta_{np}(\mathbf{k})+i\Gamma\right)\left(\omega_3-\Delta_{nm}(\mathbf{k})+i\Gamma\right)}\right.\nonumber\\[0.2cm]
    &\left. +\frac{j_{pm}^{(\mu)}(\mathbf{k})j_{mn}^{(\nu)}(\mathbf{k})j_{sp}^{(\gamma)}(\mathbf{k})j_{ns}^{(\delta)}(\mathbf{k})}{\left(\omega_1-\Delta_{ns}(\mathbf{k})+i\Gamma\right)\left(\omega_2-\Delta_{np}(\mathbf{k})+i\Gamma\right)\left(\omega_3-\Delta_{mp}(\mathbf{k})+i\Gamma\right)}\right]\text{ .}
    \label{eq:3DCS}
\end{align}
\end{widetext}

\noindent In this expression, $m,n,p$ and $s$ are band indices. Besides, the frequencies $\omega_1$, $\omega_2$ and $\omega_3$ are conjugated to the time intervals $t_1$, $t_2$ and $t_3$, respectively.

For the second-order 2DCS, on the other hand, we have contributions from single-photon vertices and two-photon vertices as we show below. As discussed in the main text, there are two types of second-order 2D spectra, called AAB and ABB spectra. 

The AAB 2DCS describes a response that is quadratic in the vector potential of the first pulse of a two-pulse sequence, and linear in the second [see Fig.\ref{fig:pulse_setup}(a)]. Therefore, both the purely paramagnetic term and the mixed of two-photon vertex and single-photon vertex in the third line of the right-hand side of Eq.(\ref{eq:K3}) compose the AAB response, while the remaining term are eliminated by the subtraction of the signal due to each pulse individually. In frequency space, the paramagnetic and the mixed AAB terms are given by 
\begin{widetext}
\begin{align}
    \chi_{\mu\nu\gamma\delta;\text{S}}^{(\text{AAB})}(\omega_1,\omega_2)=\sum\limits_{\mathbf{k},m,n,p,s}\frac{if_{ns}(\mathbf{k})}{2V}&\left[\frac{j_{nm}^{(\mu)}(\mathbf{k})j_{mp}^{(\nu)}(\mathbf{k})j_{ps}^{(\gamma)}(\mathbf{k})j_{sn}^{(\delta)}(\mathbf{k})}{\left(\omega_1-\Delta_{pn}(\mathbf{k})+i\Gamma\right)\left(\omega_2-\Delta_{mn}(\mathbf{k})+i\Gamma\right)}-\frac{j_{mp}^{(\mu)}(\mathbf{k})j_{nm}^{(\nu)}(\mathbf{k})j_{ps}^{(\gamma)}(\mathbf{k})j_{sn}^{(\delta)}(\mathbf{k})}{\left(\omega_1-\Delta_{pn}(\mathbf{k})+i\Gamma\right)\left(\omega_2-\Delta_{pm}(\mathbf{k})+i\Gamma\right)}\right .\nonumber\\[0.2cm]
    &\left.-\frac{j_{mn}^{(\mu)}(\mathbf{k})j_{pm}^{(\nu)}(\mathbf{k})j_{sp}^{(\gamma)}(\mathbf{k})j_{ns}^{(\delta)}(\mathbf{k})}{\left(\omega_1-\Delta_{np}(\mathbf{k})+i\Gamma\right)\left(\omega_2-\Delta_{nm}(\mathbf{k})+i\Gamma\right)} +\frac{j_{pm}^{(\mu)}(\mathbf{k})j_{mn}^{(\nu)}(\mathbf{k})j_{sp}^{(\gamma)}(\mathbf{k})j_{ns}^{(\delta)}(\mathbf{k})}{\left(\omega_1-\Delta_{np}(\mathbf{k})+i\Gamma\right)\left(\omega_2-\Delta_{mp}(\mathbf{k})+i\Gamma\right)}\right]\text{ ,}
    \label{eq:2DCS_aab}
\end{align}

\begin{equation}
    \chi_{\mu\nu\gamma\delta;\text{T}}^{(\text{AAB})}(\omega_1,\omega_2)
    =
    -\sum\limits_{\mathbf{k},n,m,s}
    \frac{f_{ns}(\mathbf{k})}{2V}
    \left[
    \frac{
    j_{nm}^{(\mu)}(\mathbf{k})
    j_{ms}^{(\nu)}(\mathbf{k})
    s_{sn}^{(\gamma\delta)}(\mathbf{k})
    }{
    \left(\omega_1-\Delta_{sn}(\mathbf{k})+i\Gamma\right)
    \left(\omega_2-\Delta_{mn}(\mathbf{k})+i\Gamma\right)
    }
    +
    \frac{
    j_{mn}^{(\mu)}(\mathbf{k})
    j_{sm}^{(\nu)}(\mathbf{k})
    s_{ns}^{(\gamma\delta)}(\mathbf{k})
    }{
    \left(\omega_1+\Delta_{sn}(\mathbf{k})+i\Gamma\right)
    \left(\omega_2+\Delta_{mn}(\mathbf{k})+i\Gamma\right)
    }
    \right]
    \text{ .}
    \label{eq:chi_T1}
\end{equation}

\end{widetext}

\noindent respectively. Note that $\chi_{\mu\nu\gamma\delta;\text{T}}^{(\text{AAB})}(\omega_1,\omega_2)$ has the same structure as  $\chi_{\mu\nu\gamma}^{(2)}(\omega_1,\omega_2)$ upon replacing $j_{ab}^{(\gamma)}$ with $s_{ab}^{(\gamma\delta)}$ and including an overall factor of $1/2$. 

The paramagnetic term $\chi_{\mu\nu\gamma\delta;\text{S}}^{(\text{AAB})}$ is obtained from Eq.(\ref{eq:3DCS}) through a dimensional reduction in the frequency space, $\frac{i}{\left(\omega_1-\Delta_{sn}(\mathbf{k})+i\Gamma\right)}\rightarrow \frac{1}{2}$, $\frac{i}{\left(\omega_1-\Delta_{ns}(\mathbf{k})+i\Gamma\right)}\rightarrow \frac{1}{2}$, followed by the renaming of the frequencies $\omega_2\rightarrow\omega_1$, and $\omega_3\rightarrow \omega_2$. This is equivalent to going back to the paramagnetic term of Eq.(\ref{eq:K3}), taking the limit $s_3-s_2\rightarrow 0^{+}$ and performing a Fourier transform with respect to the remaining independent time intervals. The mixed term, however, is more subtle, as one cannot just take the limit $s_3-s_2\rightarrow 0^{+}$ of the Dirac delta $\delta(s_3-s_2)$.

To handle the two-photon AAB term, we first perform the Fourier transform with respect to the three independent time intervals in $K^{(3)}$. This procedure eliminates $\delta(s_3-s_2)$ and we find Eq.(\ref{eq:chi_T1}). However, Eq.~(\ref{eq:2DCS_aab}) and Eq.(\ref{eq:chi_T1}) cannot be directly added to obtain the full AAB response, as the two reduction procedures produce quantities with different dimensions. To combine them consistently, one must return to the time-domain expression with finite width pulses. Consequently, their relative weights of the paramagnetic and mixed terms generally depend on the excitation pulse profile and cannot be uniquely determined by the ideal delta-function vector-potential pulse alone. Since specifying the temporal profile of the incident pulses is beyond the scope of this work, we analyze $\chi^{\text{(ABB)}}_\text{S}$ and $\chi^{\text{(ABB)}}_{\text{T}}$ separately.

A similar derivation holds for the ABB 2DCS, which describes the multidimensional response resulting from a linear interaction with the first pulse and quadratic with the second. It also has a purely paramagnetic contribution and a mixed two-photon vertex. The latter is given by the term on the fifth line on the right-hand side of Eq.(\ref{eq:K3}). The remaining terms do not contribute. In frequency space, they read
\begin{widetext}
\begin{align}
    \chi_{\mu\nu\gamma\delta;\text{S}}^{(\text{ABB})}(\omega_1,\omega_2)=\sum\limits_{\mathbf{k},m,n,p,s}\frac{if_{ns}(\mathbf{k})}{2V}&\left[\frac{j_{nm}^{(\mu)}(\mathbf{k})j_{mp}^{(\nu)}(\mathbf{k})j_{ps}^{(\gamma)}(\mathbf{k})j_{sn}^{(\delta)}(\mathbf{k})}{\left(\omega_1-\Delta_{sn}(\mathbf{k})+i\Gamma\right)\left(\omega_2-\Delta_{mn}(\mathbf{k})+i\Gamma\right)}-\frac{j_{mp}^{(\mu)}(\mathbf{k})j_{nm}^{(\nu)}(\mathbf{k})j_{ps}^{(\gamma)}(\mathbf{k})j_{sn}^{(\delta)}(\mathbf{k})}{\left(\omega_1-\Delta_{sn}(\mathbf{k})+i\Gamma\right)\left(\omega_2-\Delta_{pm}(\mathbf{k})+i\Gamma\right)}\right.\nonumber\\[0.2cm]
    &\left.-\frac{j_{mn}^{(\mu)}(\mathbf{k})j_{pm}^{(\nu)}(\mathbf{k})j_{sp}^{(\gamma)}(\mathbf{k})j_{ns}^{(\delta)}(\mathbf{k})}{\left(\omega_1-\Delta_{ns}(\mathbf{k})+i\Gamma\right)\left(\omega_2-\Delta_{nm}(\mathbf{k})+i\Gamma\right)} +\frac{j_{pm}^{(\mu)}(\mathbf{k})j_{mn}^{(\nu)}(\mathbf{k})j_{sp}^{(\gamma)}(\mathbf{k})j_{ns}^{(\delta)}(\mathbf{k})}{\left(\omega_1-\Delta_{ns}(\mathbf{k})+i\Gamma\right)\left(\omega_2-\Delta_{mp}(\mathbf{k})+i\Gamma\right)}\right]\text{ ,}
    \label{eq:abb}
\end{align}

\begin{equation}
    \chi_{\mu\nu\gamma\delta;\text{T}}^{(\text{ABB})}(\omega_1,\omega_2)=
    -\sum\limits_{\mathbf{k},n,m,s}
    \frac{f_{ns}(\mathbf{k})}{2V}
    \left[\frac{
    j_{nm}^{(\mu)}(\mathbf{k})
    s_{ms}^{(\nu\gamma)}(\mathbf{k})
    j_{sn}^{(\delta)}(\mathbf{k})
    }{
    \left(\omega_1-\Delta_{sn}(\mathbf{k})+i\Gamma\right)
    \left(\omega_2-\Delta_{mn}(\mathbf{k})+i\Gamma\right)
    }+
    \frac{
    j_{mn}^{(\mu)}(\mathbf{k})
    s_{sm}^{(\nu\gamma)}(\mathbf{k})
    j_{ns}^{(\delta)}(\mathbf{k})
    }{
    \left(\omega_1+\Delta_{sn}(\mathbf{k})+i\Gamma\right)
    \left(\omega_2+\Delta_{mn}(\mathbf{k})+i\Gamma\right)
    }
    \right]
    \text{ .}
\end{equation}

\end{widetext}

\noindent Note that $\chi_{\mu\nu\gamma\delta;\text{T}}^{(\text{ABB})}(\omega_1,\omega_2)$ has the same structure as  $\chi_{\mu\nu\delta}^{(2)}(\omega_1,\omega_2)$ upon replacing $j_{ab}^{(\nu)}$ with $s_{ab}^{(\nu\gamma)}$ and including an overall factor of $1/2$.  

As in the AAB response, the pure paramagnetic term is obtained by the dimensional reduction$\frac{i}{\left(\omega_2-\Delta_{pn}(\mathbf{k})+i\Gamma\right)}\rightarrow\frac{1}{2}$, $\frac{i}{\left(\omega_2-\Delta_{np}(\mathbf{k})+i\Gamma\right)}\rightarrow\frac{1}{2}$ of Eq.(\ref{eq:3DCS}), followed by a relabeling of the frequencies, while the mixed term is obtained after integrating out $\delta(s_2-s_1)$ and they are analyzed separately. It is important to note that $\chi^{(\text{ABB})}_{\text{T}}$ do not originate spectroscopic features at $\omega_2=-\omega_1$ and, therefore, do not contribute to the echo signal. 

\begin{figure*}[t!]
    \centering
\includegraphics[width=0.9\linewidth]{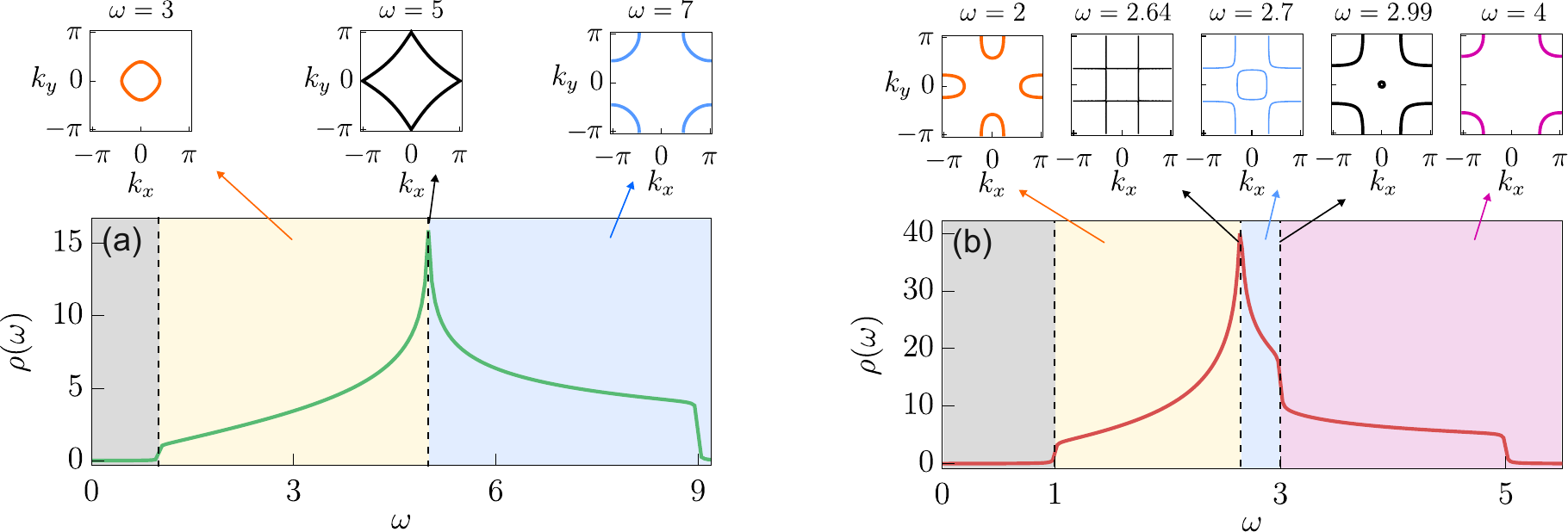}
    \caption{Joint density of states (JDOS) of a Chern insulator with (a) $M=0.5$ (topologically trivial regime) and (b) $M=-1.5$ (topologically non-trivial regime). In both panels, the gray shaded region mark the frequency range below the minimum value $\omega_{\text{min}}=1$ necessary to resonantly connect valence and conduction bands. For $\omega>\omega_{\text{min}}$, optical excitations can take place, and optically active regions (OARs) develop. The evolution of the OAR as a function of $\omega$ is shown in the insets of each panel. The vertical dashed black lines corresponds to frequencies at which the OAR changes topology, in a similar manner as a Lifshitz transitions in the Fermi surface of a metal, leading to singularities at the JDOS, which are manifested as sharp peaks in the linear and multidimensional responses.}
    \label{fig:OAR_square}
\end{figure*}

As in the $\chi^{(2)}$ case, we split the sum over the bands in $\chi^{(\text{ABB})}_{\text{S}}$ and $\chi^{(\text{ABB})}_{\text{T}}$ into the $\varepsilon_s(\mathbf{k})>\varepsilon_{n}(\mathbf{k})$ and  $\varepsilon_s(\mathbf{k})<\varepsilon_{n}(\mathbf{k})$ contributions and perform a change of variables $(n,s)\rightarrow(s,n)$ in the latter, which yields 
\begin{widetext}
\begin{align}
     \chi_{\mu\nu\gamma\delta;\lambda}^{(\text{ABB})}(\omega_1,\omega_2)=\sum\limits_{\substack{\mathbf{k},m,n,s \\ \varepsilon_s(\mathbf{k})>\varepsilon_n(\mathbf{k})}}\frac{if_{ns}(\mathbf{k})}{V}\left[L_{\mu\nu\gamma\delta;\lambda}^{(3;a)}(\omega_1,\omega_2)\right.+\left.L_{\mu\nu\gamma\delta;\lambda}^{(3;b)}(\omega_1,\omega_2)\right]\text{ ,}
\end{align}

\noindent with $\lambda=\text{S},\text{T}$, 
\begin{align}
  L_{\mu\nu\gamma\delta;\text{S}}^{(3;a)}(\omega_1,\omega_2)&=\frac{1}{2}\sum\limits_{p}\left[\frac{j_{nm}^{(\mu)}(\mathbf{k})j_{mp}^{(\nu)}(\mathbf{k})j_{ps}^{(\gamma)}(\mathbf{k})j_{sn}^{(\delta)}(\mathbf{k})}{\left(\omega_1-\Delta_{sn}(\mathbf{k})+i\Gamma\right)\left(\omega_2-\Delta_{mn}(\mathbf{k})+i\Gamma\right)}+\frac{j_{ms}^{(\mu)}(\mathbf{k})j_{pm}^{(\nu)}(\mathbf{k})j_{np}^{(\gamma)}(\mathbf{k})j_{sn}^{(\delta)}(\mathbf{k})}{\left(\omega_1-\Delta_{sn}(\mathbf{k})+i\Gamma\right)\left(\omega_2+\Delta_{ms}(\mathbf{k})+i\Gamma\right)}\right.\nonumber\\[0.2cm]
  &\left.-\frac{j_{mp}^{(\mu)}(\mathbf{k})j_{nm}^{(\nu)}(\mathbf{k})j_{ps}^{(\gamma)}(\mathbf{k})j_{sn}^{(\delta)}(\mathbf{k})}{\left(\omega_1-\Delta_{sn}(\mathbf{k})+i\Gamma\right)\left(\omega_2-\Delta_{pm}(\mathbf{k})+i\Gamma\right)}-\frac{j_{pm}^{(\mu)}(\mathbf{k})j_{ms}^{(\nu)}(\mathbf{k})j_{np}^{(\gamma)}(\mathbf{k})j_{sn}^{(\delta)}(\mathbf{k})}{\left(\omega_1-\Delta_{sn}(\mathbf{k})+i\Gamma\right)\left(\omega_2-\Delta_{mp}(\mathbf{k})+i\Gamma\right)}\right]\label{eq:L3s}\\[0.2cm]
  L_{\mu\nu\gamma\delta;\text{T}}^{(3;a)}(\omega_1,\omega_2)=&\frac{i}{2}\left[\frac{j_{nm}^{(\mu)}(\mathbf{k})s_{ms}^{(\nu\gamma)}(\mathbf{k})j_{sn}^{(\delta)}(\mathbf{k})}{
\left(\omega_1-\Delta_{sn}(\mathbf{k})+i\Gamma\right)
\left(\omega_2-\Delta_{mn}(\mathbf{k})+i\Gamma\right)
}-\frac{
j_{ms}^{(\mu)}(\mathbf{k})
s_{nm}^{(\nu\gamma)}(\mathbf{k})
j_{sn}^{(\delta)}(\mathbf{k})
}{
\left(\omega_1-\Delta_{sn}(\mathbf{k})+i\Gamma\right)
\left(\omega_2+\Delta_{ms}(\mathbf{k})+i\Gamma\right)
}\right]\text{ ,}
    \label{eq:L3}
\end{align}
\end{widetext}

\noindent and $L_{\mu\nu\gamma\delta;\lambda}^{(3;b)}(\omega_1,\omega_2)=-\left[L_{\mu\nu\gamma\delta;\lambda}^{(3;a)}(-\omega_1,-\omega_2)\right]^*$. The term $L_{\mu\nu\gamma\delta;\lambda}^{(3;a)}(\omega_1,\omega_2)$ generates resonant spectral features on the half $\omega_1\times \omega_2$ plane with $\omega_1>0$. Substituting Eq.(\ref{eq:j_Bloch}) and Eq.(\ref{eq:J1_band}) into Eq.(\ref{eq:L3s}) and Eq.(\ref{eq:L3}), we obtain $\chi_{\alpha+}^{(\text{ABB})}(\omega_1,\omega_2)$ defined in Eq.(\ref{eq:chi_ABB}), with $\alpha=\text{S},\text{T}$.

\section{Contact term contribution for the ABB response of the continuum Dirac model}\label{app:contact_Dirac}

The ABB response has both paramagnetic and contact-term contributions. In the main text, we focus on the spectroscopic signatures resulting from the paramagnetic contribution, whereas here we present the contact term separately. The contact-term contributions originate from higher-order momentum derivatives of the single-particle Hamiltonian [see Eq.(\ref{eq:chi_ABBST}) and Eq.(\ref{eq:chi_ABBT})]. For the continuum model considered in this work, these derivatives are nonzero only when $\nu=\gamma$, and therefore only the corresponding components $\chi_{\mu\nu\gamma\delta,\text{T}}^{(\text{ABB})}$ are finite. The real and imaginary parts of the $xyyx$ component for the two topological regimes are shown in Fig.\ref{fig:contact_continnum}. Its resonant energy denominator yields sharper signatures along the main diagonal and horizontal axis of the frequency plane.
%----

\section{Supplementary figures for the square lattice Chern insulator model}\label{app:JDOS_square}

In this Appendix, we provide additional details regarding the optical responses of the two-dimensional Chern insulator on the square lattice discussed in Sec.~\ref{sec:results_square}. 

\begin{figure}[t!]
    \centering
    \includegraphics[width=0.99\linewidth]{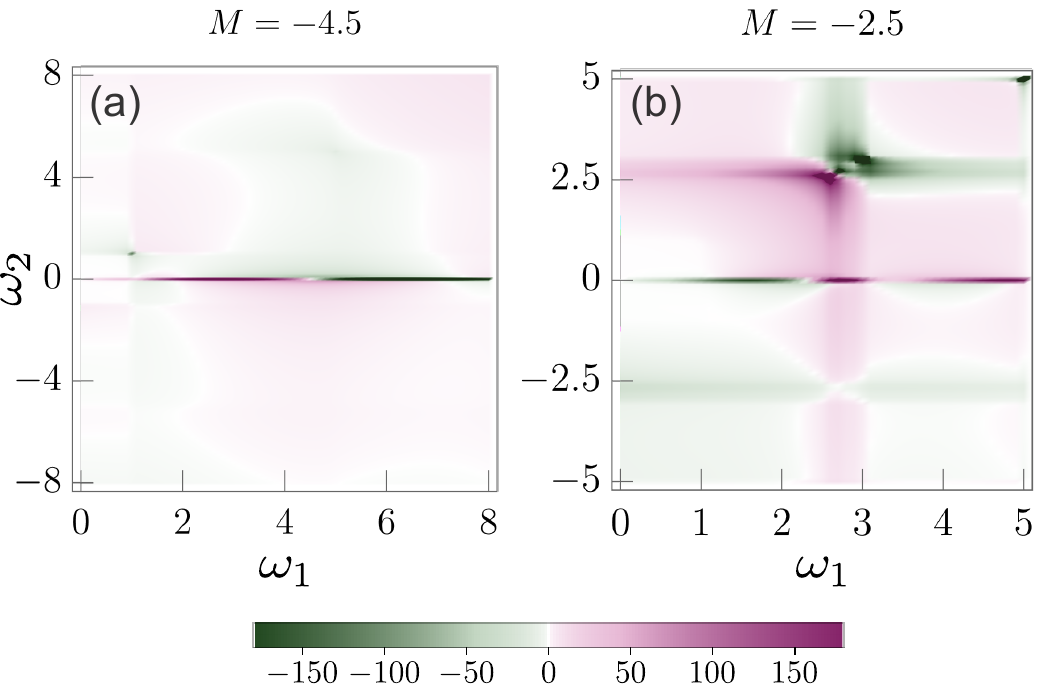}
    \caption{Contact term component of the imaginary part of the ABB $xyyx$ response (multiplied by $4\pi^2$) of the square-lattice Chern insulator with (a) $M=-4.5$ and (b) $M=-2.5$  We set $\Gamma=0.01$.}
    \label{fig:contact_square}
\end{figure}

\begin{figure}[b!]
    \centering
    \includegraphics[width=0.7\linewidth]{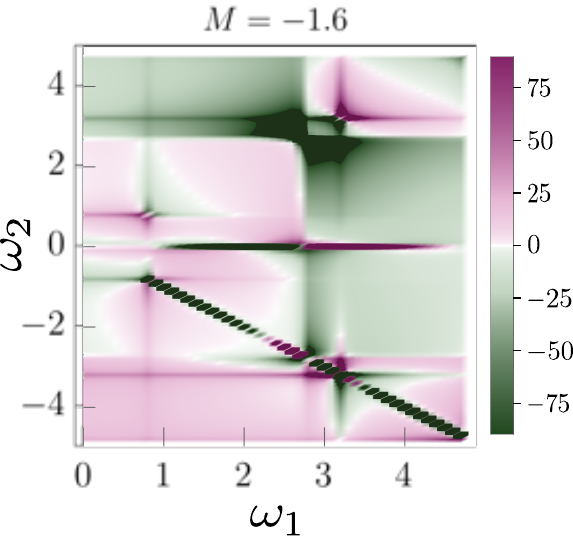}
    \caption{Imaginary part of the $xyyx$ ABB (multiplied by $4\pi^2$) response of the square-lattice Chern insulator with $M=-1.6$. We set $\Gamma=0.01$.}
    \label{fig:ABB_square_Mm1p6}
\end{figure}

We first discuss the joint density of states (JDOS) and the evolution of the optically active region (OAR) with frequency, shown in Fig.\ref{fig:OAR_square} for two values of $M$ representative of distinct topological regimes. In panel (a), $M=0.5$, corresponding to the topologically trivial regime ($C=0$). In panel (b), $M=-1.5$, corresponding to the topologically nontrivial regime with $C=-1$. The JDOS vanishes for $\omega$ below the minimum band gap, $\omega_{\text{min}}$, as indicated by the gray-shaded regions. For $\omega>\omega_{\text{min}}$, vertical optical transitions between valence and conduction bands become allowed, and the resonant condition $\Delta_{21}(\mathbf{k})=\omega$ defines the OAR at frequency $\omega$. 

The shape and connectivity of the OAR evolve with $\omega$, as shown in the insets of Fig.~\ref{fig:OAR_square}. In particular, at specific critical frequencies, the resonant contour changes its connectivity within the Brillouin zone, in a manner analogous to a Lifshitz transition in metals. At these frequencies, indicated by dashed vertical lines in Fig.~\ref{fig:OAR_square}, the JDOS exhibits singular features. The shaded regions in Fig.~\ref{fig:OAR_square} distinguish frequency intervals characterized by different OAR topologies.

Next, we consider the contact-term contributions to the ABB responses of the square-lattice Chern insulator. As in the continuum model, $\chi_{\mu\nu\gamma\delta,T}^{(ABB)}(\omega_1,\omega_2)$ is finite only for tensor components with $\nu=\gamma$. We show the imaginary part of $\chi_{xyyx,T}^{(\text{ABB})}(\omega_1,\omega_2)$ in Fig.\ref{fig:contact_square}. Since the $xyyx$ responses for values of $M$ equidistant from $M_{c,2}=-2$ are identical, we show only the spectra for $M=-4.5$ and $M=-2.5$.

Lastly, as discussed in the main text, there is no direct connection between the node-count in the echo signal of the $xyyx$ response and the band topology of the square-lattice model. In particular, the number of sign-changing nodes is even in the entire phase diagram, while its actual value can vary within the same topological phase. This is illustrated in Fig.\ref{fig:ABB_square_Mm1p6}, which shows the paramagnetic contribution to the ABB response for $M=-1.6$. In this case, the echo exhibits more nodes than for $M=-1.5$ response [see Fig.\ref{fig:2DCS_square}(c)], even though both values of $M$ belong to the same topological state.

\end{document}